\documentclass[sigconf]{sty/acmart}
\usepackage{makecell}
\usepackage{multirow}
\usepackage{xspace}
\usepackage{amstext}
\usepackage{tabularx}
\usepackage{enumitem}
\usepackage{tikz}
\usetikzlibrary{arrows.meta,positioning,calc,fit,backgrounds,patterns}
\usepackage{pgfplots}
\pgfplotsset{compat=1.18}
\usepackage[caption=false]{subfig}
\usepackage{colortbl}
\usepackage{float}
\usepackage{placeins}
\usepackage{pifont}
\usepackage[most]{tcolorbox}
\usepackage{afterpage}

\definecolor{arguscolor}{HTML}{8B4513}
\definecolor{eulercolor}{HTML}{1F77B4}
\definecolor{pikachucolor}{HTML}{2CA02C}
\definecolor{vgrnncolor}{HTML}{7B2D8E}
\definecolor{anomalecolor}{HTML}{E67E22}
\definecolor{gidslitecolor}{HTML}{E91E90}
\definecolor{baselinecolor}{HTML}{555555}
\definecolor{rred}{HTML}{D62728}
\definecolor{dkgreen}{rgb}{0,0.6,0}
\definecolor{scboxback}{HTML}{F7F9FB}
\definecolor{scboxblue}{HTML}{315F8C}
\definecolor{scboxbluealt}{HTML}{8DBBD4}
\definecolor{scboxamber}{HTML}{F1B865}
\definecolor{scboxred}{HTML}{BB3F3C}
\definecolor{scboxpurple}{HTML}{6F4BC4}
\definecolor{scboxink}{HTML}{1F2933}

\definecolor{relnovel}{HTML}{1F5FA8}
\definecolor{relinverts}{HTML}{C0392B}
\definecolor{relconfirms}{HTML}{2E7D32}
\definecolor{reladapts}{HTML}{616161}
\newcommand{\relbadge}[2]{\colorbox{#1}{\textcolor{white}{\scriptsize\bfseries\strut\makebox[36pt][c]{#2}}}}
\newcommand{\RelNovel}{\relbadge{relnovel}{Novel}}
\newcommand{\RelInverts}{\relbadge{relinverts}{Inverts}}
\newcommand{\RelConfirms}{\relbadge{relconfirms}{Confirms}}
\newcommand{\RelAdapts}{\relbadge{reladapts}{Adapts}}
\definecolor{scboxbacklight}{HTML}{EEEEEE}
\definecolor{scboxtitledark}{HTML}{4A4A4A}
\newcommand{\scbarsegment}[1]{{\color{#1}\rule{0.105\linewidth}{3pt}}}
\newcommand{\scprogressbar}[1]{%
  \begingroup
  \edef\scbarcount{\expandafter\scprogresscount#1\relax}%
  \ifnum\scbarcount>0\scbarsegment{anomalecolor}\fi
  \ifnum\scbarcount>1\scbarsegment{pikachucolor!55}\fi
  \ifnum\scbarcount>2\scbarsegment{scboxamber}\fi
  \ifnum\scbarcount>3\scbarsegment{scboxbluealt}\fi
  \ifnum\scbarcount>4\scbarsegment{vgrnncolor!45}\fi
  \endgroup
}
\def\scprogresscount#1/#2 systems\relax{#1}

\newcommand{\sysdot}[1]{\protect\tikz[baseline=-0.55ex]\protect\fill[#1] (0,0) circle (0.075);\,}

\DeclareRobustCommand{\Argus}{\textsc{Argus}\xspace}
\DeclareRobustCommand{\Euler}{\textsc{Euler}\xspace}
\DeclareRobustCommand{\Pikachu}{\textsc{Pikachu}\xspace}
\DeclareRobustCommand{\Vgrnn}{\textsc{VGRNN}\xspace}
\DeclareRobustCommand{\AnomalE}{\textsc{Anomal-E}\xspace}
\DeclareRobustCommand{\GidsLite}{\textsc{GIDS-Lite}\xspace}

\DeclareRobustCommand{\Jbeil}{\textsc{Jbeil}\xspace}
\DeclareRobustCommand{\EGraphSAGE}{\textsc{E-GraphSAGE}\xspace}
\DeclareRobustCommand{\Arganids}{\textsc{ARGANIDS}\xspace}

\DeclareRobustCommand{\ArgusM}{\sysdot{arguscolor}\textsc{Argus}\xspace}
\DeclareRobustCommand{\EulerM}{\sysdot{eulercolor}\textsc{Euler}\xspace}
\DeclareRobustCommand{\PikachuM}{\sysdot{pikachucolor}\textsc{Pikachu}\xspace}
\DeclareRobustCommand{\VgrnnM}{\sysdot{vgrnncolor}\textsc{VGRNN}\xspace}
\DeclareRobustCommand{\AnomalEM}{\sysdot{anomalecolor}\textsc{Anomal-E}\xspace}
\DeclareRobustCommand{\GidsLiteM}{\sysdot{gidslitecolor}\textsc{GIDS-Lite}\xspace}
\DeclareRobustCommand{\NetWalkM}{\sysdot{baselinecolor}\textsc{NetWalk}\xspace}
\DeclareRobustCommand{\JbeilM}{\sysdot{baselinecolor}\textsc{Jbeil}\xspace}
\DeclareRobustCommand{\EGraphSAGEM}{\sysdot{baselinecolor}\textsc{E-GraphSAGE}\xspace}
\DeclareRobustCommand{\ArganidsM}{\sysdot{baselinecolor}\textsc{ARGANIDS}\xspace}

\DeclareRobustCommand{\VgrnnPlain}{\textsc{VGRNN}\xspace}

\DeclareRobustCommand{\GidsLitePlain}{\textsc{GIDS-Lite}\xspace}

\newtcolorbox{scbox}[3]{
  enhanced,
  arc=2pt,
  boxrule=0.5pt,
  colback=scboxback,
  colframe=scboxblue,
  left=6pt,
  right=6pt,
  top=3.5pt,
  bottom=3.5pt,
  before upper={\noindent{\bfseries\color{scboxink}#1}\par
    \vspace{2.5pt}\noindent
    {\color{scboxtitledark!35}\rule[0.5ex]{\dimexpr\linewidth-5.8em\relax}{0.5pt}}%
    \hfill{\footnotesize\itshape\color{scboxtitledark}#3}\par
    \vspace{3.5pt}\noindent #2\par}
}

\newcommand{\cmark}{{\color{dkgreen}\ding{51}}}
\newcommand{\xmark}{{\color{rred}\ding{55}}}

\newlength\tbspace
\newcolumntype{C}{c<{\hspace{\tbspace}}}

\newcommand{\etal}{\textit{et al}.\xspace}

\newcommand{\Sys}{GIDS-Eval\xspace}
\newcommand{\gnn}{graph neural network\xspace}

\newcommand{\gids}{GIDS\xspace}

\newcommand{\optc}{OpTC\xspace}
\newcommand{\lanl}{LANL\xspace}
\newcommand{\cicids}{CI-2017\xspace}

\newcommand{\ap}{AP\xspace}
\newcommand{\auc}{AUC\xspace}
\newcommand{\mcc}{MCC\xspace}

\newcommand{\cdr}{CDR\xspace}
\newcommand{\adp}{ADP\xspace}
\newcommand{\tpr}{TPR\xspace}
\newcommand{\fpr}{FPR\xspace}

\newcommand{\numconsidered}{nine\xspace}
\newcommand{\numsystems}{five\xspace}
\newcommand{\numgaps}{nine\xspace}
\newcommand{\numdatasets}{four\xspace}
\newcommand{\numconfigs}{240\xspace}
\newcommand{\numablation}{1{,}370\xspace}%

\copyrightyear{2026}
\acmYear{2026}
\setcopyright{cc}
\setcctype{by}
\acmConference[CCS '26]{Proceedings of the 2026 ACM SIGSAC Conference on Computer and Communications Security}{November 15--19, 2026}{The Hague, Netherlands}
\acmBooktitle{Proceedings of the 2026 ACM SIGSAC Conference on Computer and Communications Security (CCS '26), November 15--19, 2026, The Hague, Netherlands}
\acmDOI{10.1145/3830454.3846576}
\acmISBN{979-8-4007-2871-6/2026/11}
\acmBadge[https://www.acm.org/publications/policies/artifact-review-and-badging-current]{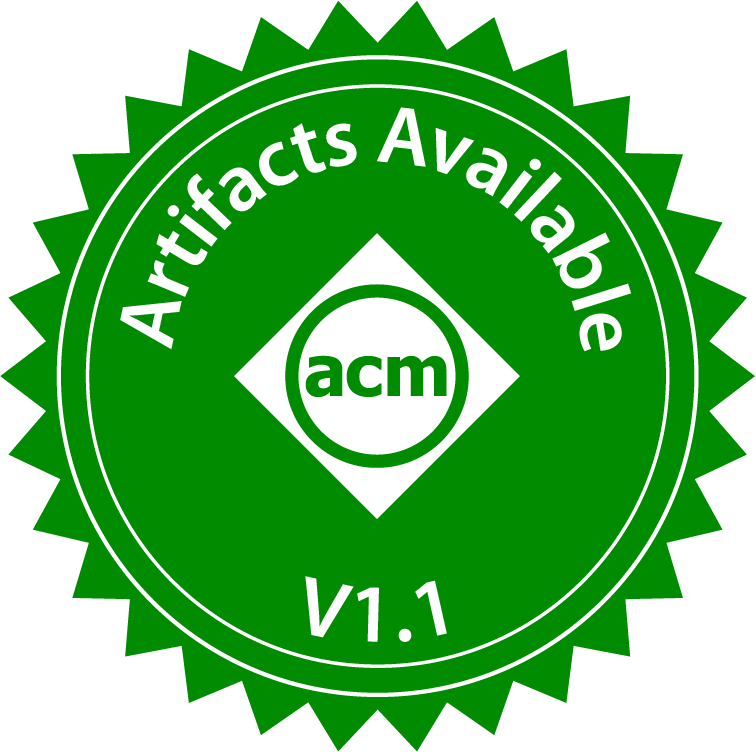}
\acmBadge[https://www.acm.org/publications/policies/artifact-review-and-badging-current]{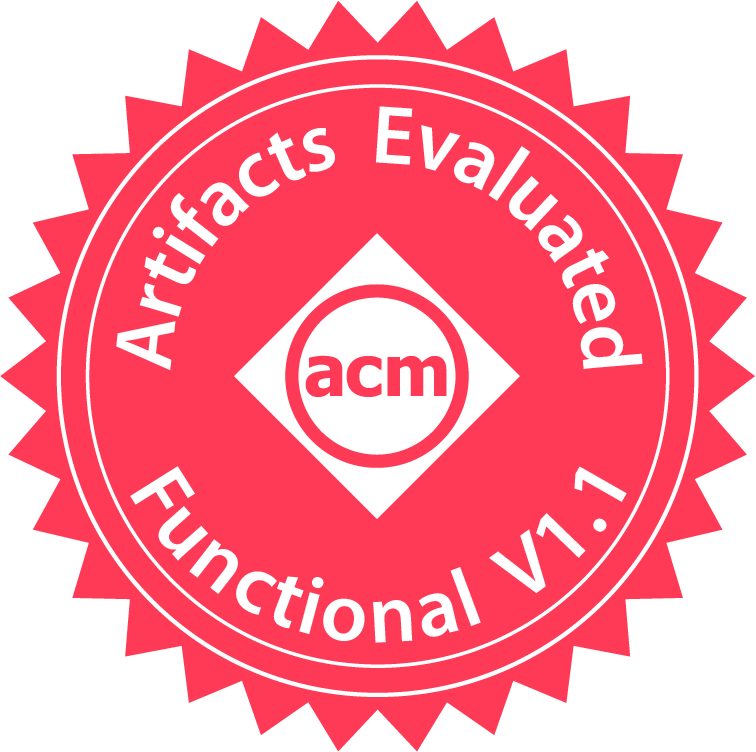}
\acmBadge[https://www.acm.org/publications/policies/artifact-review-and-badging-current]{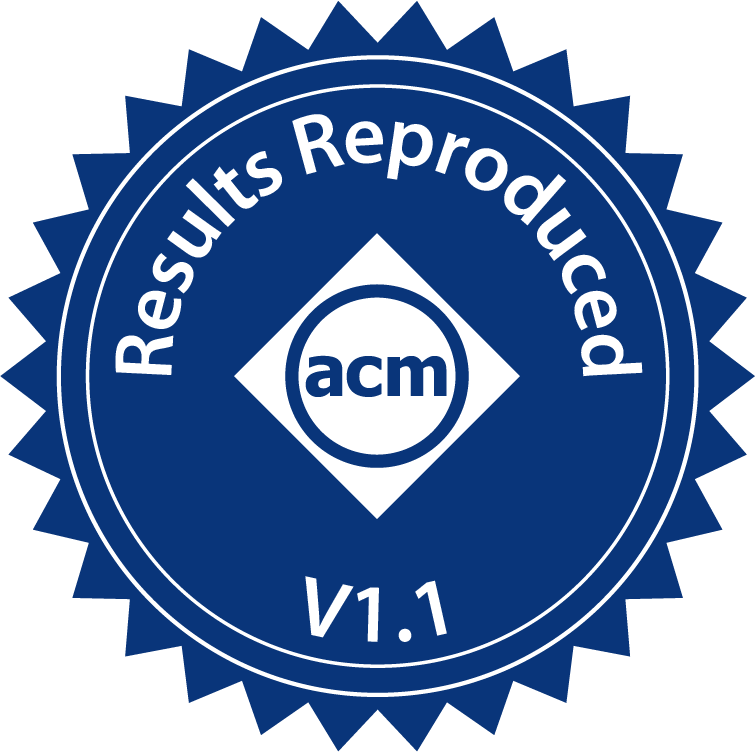}
\author{Rui Zhao}
\affiliation{%
  \institution{University of Virginia}
  \city{Charlottesville}
  \country{USA}
}
\email{dkw7xn@virginia.edu}

\author{Wajih Ul Hassan}
\affiliation{%
  \institution{University of Virginia}
  \city{Charlottesville}
  \country{USA}
}
\email{hassan@virginia.edu}

\begin{document}

\title{A First-Principles Evaluation of Graph-Based Network Intrusion
Detection Systems}

\begin{abstract}
Graph-based network intrusion detection systems (\gids) report strong
benchmark detection metrics, but those metrics establish little about
deployability.  We approach the problem from first principles: rather
than inheriting the preprocessing, windowing, and thresholding
conventions of each published system, we ask what a controlled
comparison requires and impose it uniformly.  The result is \Sys, an
evaluation framework that decomposes a \gids into six interchangeable
stages and turns those conventions into explicit experimental
variables, so reported performance can be attributed to
individual stages instead of whole pipelines.  We survey \numconsidered
representative \gids, reimplement \numsystems of them within \Sys, and
evaluate them on \numdatasets datasets under one matched protocol.  We
identify \numgaps recurring evaluation gaps and quantify the impact of
each: two crafted edges achieve full evasion against three of the
eight detector--dataset pairs with anything to hide; the snapshot
window alone accounts for a mean 38.3\%
relative swing in average precision (\ap); aligning preprocessing
across systems moves \ap by up to 61.8 percentage points for a single
detector; and none of the 18 detector--dataset pairs we replay can
alert as events arrive.  We introduce \GidsLitePlain,
an encoder-free control built in the same framework, which ranks
first by \ap on two of the \numdatasets datasets at up to
575$\times$ lower runtime.  Architectural complexity is therefore not
a consistent driver of detection quality under our matched protocol on
current benchmarks, but it does enlarge the runtime, calibration, and
attack surfaces operators must defend.
\end{abstract}



\maketitle

\section{Introduction}
\label{s:intro}

Network intrusion detection remains a core defense for enterprise
security, particularly against attacks that unfold across many network
events rather than a single malicious
connection~\cite{uetz2024siem,wei2023xnids,piet2023ssh}.  Graph-based
Network Intrusion Detection Systems (\gids) address this setting by
applying graph neural networks and temporal models to network traffic,
aiming to detect attacker behaviors such as lateral movement across
hosts~\cite{khoury2024jbeil}.  Recent \gids~\cite{argus2020,
king2022euler,lo2022graphsage,pikachu2023,vgrnn2019,anomale2022,
khoury2024jbeil} combine graph encoders such as GCNs~\cite{kipf2017semi}
and GraphSAGE~\cite{hamilton2017inductive} with temporal modules such
as GRUs~\cite{cho2014learning} and recurrent graph
models~\cite{vgrnn2019}, and typically report very high headline
detection metrics on public datasets, even as several acknowledge
precision limitations.  From a security standpoint, however, benchmark
accuracy is not deployability.  The architectural choices that produce
a strong precision number on a fixed snapshot also define a
graph-scoring path an attacker can target: in our experiments, two
crafted edges inserted into the network graph hide every attack edge
that state-of-the-art \gids catch, while leaving every
other detector input unchanged.  The relevant
security question is not whether a \gids detector wins a clean benchmark,
but whether its decisions survive an attacker who can shape the graph it
reads, an operator who must wire it into a streaming sensor, and the
pipeline choices that current \gids publications leave undocumented.

\begin{table*}[!t]
\centering
\begin{footnotesize}
\setlength{\tabcolsep}{5pt}
\renewcommand{\arraystretch}{1.15}
\begin{tabular}{@{}l p{0.27\textwidth} p{0.50\textwidth}@{}}
\toprule
\textbf{System}
  & \textbf{Graph Setting}
  & \textbf{Core Technique} \\
\midrule
\multicolumn{3}{@{}l}{\textit{Implemented in our modular framework}} \\
\ArgusM~\cite{argus2020}
  & Discrete temporal graph over enterprise logs
  & MPNN encoder with edge features, attack-aware decoder, and AP-oriented loss \\
\EulerM~\cite{king2022euler}
  & Discrete temporal host graph
  & Graph encoder plus sequence encoder for temporal link prediction \\
\PikachuM~\cite{pikachu2023}
  & Dynamic graph over time-ordered traffic events
  & Temporal walks, Skip-gram embeddings, and GRU autoencoder \\
\VgrnnM~\cite{vgrnn2019}\textsuperscript{\dag}
  & Dynamic graph snapshots
  & Variational graph recurrent model for latent node embeddings and link reconstruction \\
\AnomalEM~\cite{anomale2022}
  & Flow graph with edge attributes
  & E-GraphSAGE and modified DGI with a classical anomaly detector \\
\midrule
\multicolumn{3}{@{}l}{\textit{Surveyed but not implemented}} \\
\NetWalkM~\cite{netwalk}\textsuperscript{a}
  & Dynamic network stream
  & Clique-based network embedding with deep autoencoder regularization and clustering \\
\JbeilM~\cite{khoury2024jbeil}\textsuperscript{b}
  & Continuous-time authentication graph
  & TGN-style memory encoder and decoder for inductive lateral-movement link prediction \\
\EGraphSAGEM~\cite{lo2022graphsage}\textsuperscript{c}
  & Flow graph with edge attributes
  & Edge-aware GraphSAGE for supervised intrusion detection on flow-based traffic graphs \\
\ArganidsM~\cite{arga}\textsuperscript{d}
  & Flow graph with node features
  & Adversarially regularized graph autoencoder for unsupervised intrusion detection \\
\bottomrule
\end{tabular}
\caption{Representative \gids considered in this study. The upper block lists
the \numsystems systems implemented in \Sys. The lower block lists additional
surveyed systems that define the broader design space but are excluded from
full modular evaluation for the reasons in the footnotes.}
\label{tab:gids-overview}
\begin{minipage}{0.96\textwidth}
\scriptsize
\textsuperscript{\dag} A general dynamic-graph model rather than a \gids,
studied as the temporal graph encoder backbone later \gids adopted for
intrusion detection.
\textsuperscript{a} Runtime is impractical for full system-dataset ablations.
\textsuperscript{b} The released implementation leaves the paper's
preprocessing and evaluation protocol underspecified and omits details needed
for faithful modularization.
\textsuperscript{c} The supervised design requires extensive labeled training
data, outside our evaluation protocol.
\textsuperscript{d} No public implementation is available, and the paper lacks
the detail needed for faithful modularization.
\end{minipage}
\end{footnotesize}
\end{table*}

This deployability gap has three practical consequences.  First, complex
graph and temporal designs enlarge the attack surface, as the
covering-edge result above demonstrates.  Second, they increase
training time, memory demand, and tuning burden, raising operational
cost.  Third, existing \gids results do not establish what actually
drives the reported performance, because each paper bundles a different
preprocessing pipeline, snapshot window, feature representation,
thresholding procedure, and metric choice.  Performance differences
across papers therefore cannot be attributed to the graph or temporal
architecture itself.  An operator deploying a recent \gids detector
inherits all three problems and cannot tell whether the added complexity
yields any defensive benefit.

Recent evidence suggests this concern is not theoretical.  The R+R
study by Wang \etal~\cite{wang2025rr} re-ran five published \gids
end to end and showed that reported results vary substantially across
preprocessing choices, dataset splits, and evaluation protocols,
treating each system as a black box.  In parallel, Bilot
\etal~\cite{bilot2025simpler} showed that, in provenance-based
intrusion detection (PIDS), a simple neural network matches or
outperforms much more complex graph-based systems under a unified
evaluation, and identified nine PIDS evaluation shortcomings.  Neither
study establishes whether the architectural complexity of modern \gids
provides a real defensive advantage on the axes that matter to a security
operator: adversarial robustness, calibration stability under streaming
traffic, and detection latency.  We answer that question directly.

To do so, we build \Sys, a modular evaluation framework that decomposes
each \gids into six interchangeable stages: feature extraction, graph
transformation, featurization, encoding, decoding, and optimization.
We survey \numconsidered representative \gids and
implement \numsystems within \Sys (Table~\ref{tab:gids-overview}),
running \numablation experimental jobs in total: 960 of them are the
\numconfigs-configuration component grid of EG7 run on each of the
\numdatasets datasets, and the rest are the per-gap sweeps of
\S\ref{s:shortcomings}.  The decomposition turns non-architectural
choices such as snapshot windowing and threshold calibration into
explicit experimental variables, so performance differences can be attributed to
specific stages rather than whole pipelines.  Within \Sys, we introduce
\GidsLitePlain, a deliberately simple control that scores each
network event using only featurization and a shallow predictor,
without a graph or temporal encoder.  \GidsLite is not another
specialized architecture.  It is a control that tests under a matched
protocol whether the complexity of recent \gids yields a detection or
robustness advantage that justifies its cost.

Through \Sys, we identify \numgaps recurring evaluation gaps that
limit the practical security value of current \gids.  The most
security-critical is the adversarial fragility above, which a
gradient-guided covering-edge attack~\cite{xu2023cover} exposes (EG6).
The remaining gaps show that even before an adversary intervenes,
reported performance is dominated by non-architectural pipeline
choices a real deployment cannot avoid: snapshot-window choice alone
swings average precision (\ap) by a mean of 38.3\% across systems and
datasets (EG1), aligning the preprocessing pipeline across systems
shifts \ap by up to 61.8 percentage points for a single
detector--dataset pair (EG8), and none of the 18 detector--dataset
pairs we replay can alert as events arrive (EG9).  The others, from
seed- and test-period-dependent thresholding to edge attributes that
are near-collinear across attack and benign traffic, are quantified in
\S\ref{s:shortcomings}.  Each gap affects the security guarantees a
\gids can offer in deployment, and most are not examined in the
original publications.

These findings extend prior work in two directions.  R+R established
\emph{that} \gids pipeline choices matter while treating each system as
a black box.  \Sys attributes the variance to specific stages,
quantifies each as a single axis with the numbers above, and uses
\GidsLite as a control R+R does not propose.  The PIDS results of
Bilot \etal do not transfer cleanly because the data models differ: provenance graphs
are given by kernel audit causality~\cite{inam2022sok}, labeled under
converged DARPA Transparent Computing (TC) conventions~\cite{darpa},
and described by symbolic
features such as file paths and command
lines~\cite{flash2024}, whereas \gids must build the graph from a
continuous flow stream, have no canonical preprocessing contract, and
rely on statistical edge attributes that look almost identical for
attack and benign traffic.  Of the \numgaps gaps we identify, two are
measurements neither prior study provides (EG1, EG8).  Three sit on
axes Bilot \etal also examine, with the answer inverted on network flows
(EG4, EG6, EG7).  Three sharpen shared evaluation-hygiene concerns into
operator-facing measurements (EG2, EG3, EG5).  One, streaming
readiness (EG9), carries the deployment axis Bilot \etal address for PIDS
to \gids with a replay measurement.
This paper thus extends R+R from black-box reproduction to
stage-level attribution, and extends Bilot \etal from PIDS to a network
setting whose data model creates gaps PIDS evaluation cannot
encounter.

Under the matched protocol of \Sys, one fixed \GidsLite configuration
ranks first by \ap on two of the four datasets while reaching up
to $\sim$575$\times$ lower end-to-end runtime than the slowest system
and avoiding the
graph-scoring path the covering-edge attack exploits.  On the other
two datasets the winners are different architectures, \Pikachu and
\Vgrnn.  Taken together
with the gap results, this shows that the architectural
complexity of recent \gids is not a consistent driver of detection
quality on current benchmarks, while it does enlarge the runtime,
calibration, and adversarial attack surfaces an operator must defend.

We make the following contributions:

\begin{table*}[!t]
\centering
\begin{footnotesize}
\setlength{\tabcolsep}{5pt}
\renewcommand{\arraystretch}{1.1}
\begin{tabular}{@{}l l l l l l l@{}}
\toprule
\textbf{System}
  & \textbf{Feature Extraction}
  & \textbf{Graph Transform.}
  & \textbf{Featurization}
  & \textbf{Encoding}
  & \textbf{Decoding}
  & \textbf{Optimization} \\
\midrule
\ArgusM~\cite{argus2020}
  & src/dst IP, port, flow stats
  & Temporal snapshots
  & Flow features
  & MPNN + GRU
  & Edge scoring
  & Avg.\ precision loss \\
\EulerM~\cite{king2022euler}
  & src/dst host identifiers
  & Temporal snapshots
  & Identity emb.
  & GCN + GRU
  & Adj.\ reconstruction
  & Reconstruction loss \\
\PikachuM~\cite{pikachu2023}
  & src/dst IP, port, protocol
  & Temporal snapshots
  & Random walk emb.
  & Skip-gram + GRU
  & Link prediction
  & Binary cross-entropy \\
\VgrnnM~\cite{vgrnn2019}
  & src/dst host identifiers
  & Temporal snapshots
  & Identity emb.
  & GCN + GC-LSTM
  & Adj.\ reconstruction
  & Recon.\ + KL-div.\ loss \\
\AnomalEM~\cite{anomale2022}
  & src/dst IP, flow statistics
  & Static graph
  & Flow features
  & GraphSAGE
  & Node feat.\ prediction
  & Node prediction loss \\
\bottomrule
\end{tabular}
\caption{Stage-level decomposition of the \numsystems systems implemented in
\Sys. Unlike Table~\ref{tab:gids-overview}, which defines the broader study
scope, this table maps only the implemented systems to the six stages exposed
by the modular framework. Encoding lists the graph encoder and, where
applicable, the temporal encoder separated by ``+''. emb. = embedding. Recon. = reconstruction. }
\label{tab:systemization}

\end{footnotesize}
\end{table*}

\begin{itemize}[leftmargin=*,topsep=1pt,itemsep=1pt,parsep=0pt]

\item We present \Sys, a modular evaluation framework that maps
representative systems into six shared, interchangeable stages, and
run \numablation jobs, among them a
\numconfigs-configuration component grid, to attribute reported \gids
performance to specific stages rather than whole pipelines.

\item We identify \numgaps recurring evaluation gaps that limit the
practical security value of current \gids, and classify each against
the two prior studies as novel, inverted, confirmed, or adapted.

\item We show, via a gradient-guided covering-edge attack, that two
inserted edges hide every attack edge state-of-the-art \gids
catch, and that six of the eight detector--dataset pairs with anything
to hide fall within twenty, exposing a graph-side attack surface their
architectural complexity creates rather than mitigates.

\item We introduce \GidsLitePlain, a first-principles baseline that,
in a single configuration fixed on validation data, ranks first by \ap
on half the datasets at a fraction of the runtime.

\end{itemize}

\paragraph{Availability.}
\begingroup\sloppy
The source code and paper artifact are available at
\url{https://github.com/DART-Laboratory/GIDS-Eval}.
\endgroup

\section{Background \& Study Scope}\label{s:background}

This section summarizes the design-space progression of recent \gids
and the systems we study. \Sys is introduced in \S\ref{s:framework}.

\subsection{Evolving \gids Complexity}
\label{s:gids-complexity}

Recent \gids occupy a spectrum of design choices.  At the simpler end,
systems such as \EGraphSAGE~\cite{lo2022graphsage},
\AnomalE~\cite{anomale2022}, and \Arganids~\cite{arga} construct a traffic
graph, apply a graph encoder or reconstruction objective, and treat
classification, reconstruction error, or representation quality as the
anomaly signal.  The modeling assumption is that malicious activity
distorts local graph structure or endpoint neighborhoods in a way the
encoder can pick up.  Later designs add temporal modeling on top of
this graph abstraction: \Euler~\cite{king2022euler} predicts links over
temporal host graphs, \Pikachu~\cite{pikachu2023} combines temporal
walks with a GRU autoencoder, and \Vgrnn~\cite{vgrnn2019} uses
variational graph recurrent dynamics over evolving snapshots, so the
anomaly score depends jointly on graph construction, temporal
windowing, sequence modeling, and the prediction target.  At the most
coupled end, \Argus~\cite{argus2020} combines
message passing over temporal enterprise graphs with edge features, an
attack-aware decoder, and an AP-oriented loss, and
\Jbeil~\cite{khoury2024jbeil} uses memory-based temporal graph modeling
for inductive lateral-movement link prediction.  Here every stage
shapes the anomaly signal together.

\subsection{Representative Systems}\label{s:systems}

Table~\ref{tab:gids-overview} lists the \numconsidered representative
\gids in our survey, spanning flow and host graphs, static and temporal
graph construction, graph and temporal encoders, reconstruction and
prediction objectives, and separate anomaly scoring components.  From
this set, we select \numsystems for in-depth evaluation.  Three criteria governed
inclusion: the system targets network-level intrusion detection, its
design appeared at a peer-reviewed venue, and its release is complete
enough, as runnable code or a fully specified description, to rebuild
faithfully from interchangeable modules.  \Vgrnn meets the first
criterion indirectly.  It was published as a variational graph
recurrent model, not as an intrusion detector, and we evaluate the
intrusion-detection instantiation that later \gids work built on
it~\cite{king2022euler,wang2025rr}.  In this paper it is a
temporal graph encoder backbone, not a purpose-built \gids.
The remaining
systems stay in the survey scope but are excluded from full evaluation
for insufficient public implementation detail, supervision
requirements outside our protocol, or scale constraints.  Faithful
modularization matters because each implemented system is decomposed
into independent components that \Sys re-combines and varies across
configurations.

\section{The \Sys Framework and Protocol}
\label{s:framework}

\begin{figure*}[!t]
\centering
\includegraphics[width=\textwidth]{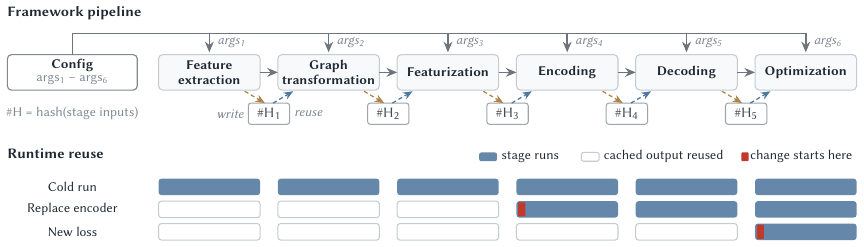}
\Description{The GIDS-Eval pipeline. A configuration file feeds one
  argument block to each of six stages arranged left to right. Between
  consecutive stages, outputs are written to and reused from
  content-addressed caches. Below, three runtime scenarios show which
  stages run and which are served from the cache when a configuration
  changes.}
\caption{The \Sys pipeline.  A configuration supplies one argument
  block per stage.  Every stage but the last writes
  its output to a cache keyed by a hash of its inputs, and a
  stage with unchanged inputs is served from that cache: replacing
  the encoder re-runs only the encoding, decoding, and optimization
  stages, and changing only the loss re-runs the final stage
  alone.}
\label{fig:pipeline}
\end{figure*}

\paragraph{Modular design.}
\Sys starts from what a controlled comparison requires instead of
adopting any single system's preprocessing, windowing, and
thresholding conventions, and imposes those requirements uniformly.
The staged decomposition that makes this
possible follows the approach Bilot \etal~\cite{bilot2025simpler}
introduced for provenance-based systems, adapted here to the
network-flow setting.
To vary one pipeline stage while holding the rest fixed, we
integrate the \numsystems studied detectors as interchangeable
configurations of \Sys, whose shared infrastructure builds on PyTorch
and PyTorch Geometric~\cite{pyg}.  The columns of
Table~\ref{tab:systemization}, which follows the form of their PIDS
systemization with the stages redefined for network flows, define the
six stages every integrated system shares.  Raw network records enter at
\textbf{feature extraction}, which selects the fields that populate
the graph.  \textbf{Graph transformation} builds a static or temporally
snapshotted graph from those records.  \textbf{Featurization} attaches
numeric attributes to nodes and edges.  \textbf{Encoding} learns vector
representations with a graph encoder and, optionally, a temporal
encoder.  \textbf{Decoding} maps the representations to a per-edge,
per-node, or per-graph anomaly signal.  Finally, \textbf{optimization}
supplies the loss that trains the encoder and decoder.  Each cell in
Table~\ref{tab:systemization} corresponds to a module, and modules
within the same stage are interchangeable.

A single YAML file selects, for each stage, one module and its
hyperparameters, so every experiment is fully described and
reproducible from that file (Figure~\ref{fig:pipeline}).  \Sys hashes
each stage's inputs and caches the outputs of every stage but the
last in content-addressed folders, so a stage whose inputs are unchanged is skipped and its
cached output feeds directly to the next stage.  Replacing a graph
encoder, for example, invalidates only the encoding, decoding, and
optimization stages, while earlier stages are reused across configurations
that share the same upstream choices.  This caching is what makes the
\numconfigs-configuration component-attribution sweep in EG7 tractable
across Campus, CI-2017, \lanl, and \optc.

\paragraph{Datasets.}
We evaluate on \numdatasets datasets that span distinct deployment
environments, telemetry types, scales, and attack-labeling mechanisms.
Campus is a large-scale enterprise gateway trace introduced in prior
R+R work~\cite{wang2025rr}.  \lanl~\cite{lanl} provides authentication
and flow telemetry with human red-team activity, the setting in
which Active Directory attacks unfold~\cite{hades2025}.
\optc~\cite{optc} provides
mixed host and network telemetry from a simulated enterprise
engagement, and CI-2017, the corrected-label variant~\cite{liu2022error}
of CIC-IDS-2017~\cite{cicids}, provides flow-level records with scripted
attacks in a controlled lab environment.  We preserve temporal order
when constructing train, validation, and test splits.
Appendix~\ref{app:datasets} reports the per-dataset statistics.

The \numdatasets datasets differ in fidelity.  Campus is the weakest
case: its attack traffic is injected as scripted APT
campaigns~\cite{wang2025rr}, so its attack distribution need not match
how those attacks appear in production, and it is one of the two
datasets on which no evaluated \gids reaches a deployable operating
point.  \optc
is likewise generated in a simulated enterprise and carries synthesis
artifacts of its own.  For CI-2017 we use the
corrected-label variant~\cite{liu2022error} rather than the original
release, avoiding the labeling and construction problems documented
for benchmark NIDS datasets~\cite{flood2024smells}.  We keep all
\numdatasets datasets, and all \numdatasets are publicly available:
their contrasts across telemetry type, scale, and labeling are what
make a cross-dataset finding more than a property of one benchmark.
No single dataset should be read as a proxy for deployment.

\paragraph{Metrics.}
We measure detector behavior along four dimensions: score ranking,
thresholded detection, campaign coverage, and execution cost.  For
score ranking, we report \ap and \auc before any threshold is chosen,
treating \ap, the area under the precision--recall curve, as the
primary threshold-free metric because attack events are rare and
ROC-style metrics can hide false-alert burden~\cite{sommer2010outside}.  For
thresholded detection, we select thresholds on the validation split, compute TP, FP,
FN, and TN, and report \tpr, \fpr, precision, recall, and $F_1$.  For
execution cost, we record training time, inference time, wall-clock
time, peak memory, and CPU utilization.

Event-level metrics do not capture whether a detector exposes a
multi-stage intrusion campaign: a system can score well in aggregate
while missing every step of an attack, or expose a campaign with a
single high-confidence alert.  Motivated by the \adp metric of Bilot
\etal~\cite{bilot2025simpler}, we introduce \cdr for \gids evaluation.
Let $\mathcal{C}$ be the set of attack campaigns in the evaluated test
period and
$\hat{\mathcal{C}}$ the campaigns for which the detector flags at least
one positive event.  We define \cdr as
$|\hat{\mathcal{C}}|/|\mathcal{C}|$.

\paragraph{Hardware setup.}
We ran all experiments on virtual machines hosted on a dual-socket AMD
EPYC 9654 server with 192 physical cores, 384 hardware threads, and
800\,GB RAM, running Ubuntu 22.04 with Linux kernel 6.5.0.
Each VM had 36 vCPUs and 128\,GB RAM and only
the software required for the evaluated system.  Datasets were stored
on the host and accessed through VFS, isolating each software
environment on shared hardware.

\paragraph{Implementation and validation.}
All \numsystems studied systems provide public source code.  We start
from the released implementations of \Argus, \Euler, \Pikachu, \Vgrnn,
and \AnomalE, and preserve their model architectures, preprocessing,
training objectives, hyperparameters, and scoring logic.  Our changes
are limited to wrapping each codebase behind the six-stage module
interface above: the wrappers translate input and output formats, set
dataset paths, and collect runtime logs, but do not replace the
detector-specific learning code.  We validate each wrapper by running
the original code on the datasets and settings supported by its release
and checking that the wrapper produces consistent results, using the
configuration reported in the corresponding paper when a release
exposes multiple settings.  For experiments that test a specific design
choice, such as the snapshot window, we change only that module while keeping
the rest of the released pipeline fixed.

\section{Evaluation Gaps}\label{s:shortcomings}
We use \emph{evaluation gap} for an aspect of a \gids that current
evaluations fail to isolate or report, and each of the \numgaps gaps
below is stated in that form.  Four are failures to \textbf{isolate} a
design choice from the result it influences: the snapshot window
(EG1), the edge-feature path (EG4), the individual pipeline components
(EG7), and the preprocessing pipeline (EG8).  One is a failure to
\textbf{separate} calibration from test evaluation (EG3), which is
also a design choice of the one system that uses it.  The
remaining four are failures to \textbf{measure or report} a dimension
on which a deployed detector is judged: threshold variance (EG2),
resource cost (EG5), adversarial robustness (EG6), and event-to-alert
latency (EG9).  The isolate and separate gaps concern experimental
control, the measure-or-report gaps concern reporting coverage, and a
system carries a mark in Table~\ref{tab:sc-summary} when its own
evaluation leaves that gap open, because its
design confounds the dimension or because it never measured it.

The \numgaps gaps examined below emerged from our
component-isolation experiments on the \numsystems systems of
\S\ref{s:systems}, and \S\ref{s:eval} provides the complementary
end-to-end comparison under the shared protocol.  Space permits only
the headline result for each gap, and Appendix~\ref{app:per-sc}
reports the complete numbers.

\begin{scbox}{EG1: Snapshot-Window Sensitivity}{%
The snapshot window used to group event streams into graphs is a hidden
evaluation variable.  Holding the detector and dataset fixed while varying
only this window produces more than 20\% relative \ap swing in 11 of 16
detector--dataset pairs, showing that reported accuracy can depend heavily
on an under-documented preprocessing choice.}{4/5 systems}
\end{scbox}

\begin{figure*}[!t]
\centering
\captionsetup{skip=1pt}
\includegraphics[width=\textwidth]{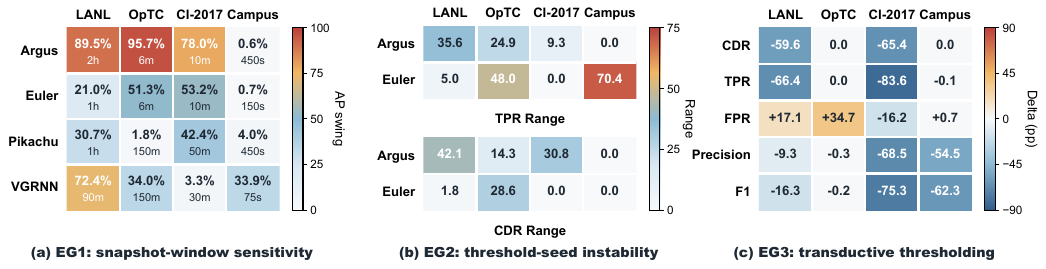}
\Description{Three aligned heatmap panels. Panel a shows snapshot-window
  sensitivity for EG1, panel b shows threshold-seed instability for EG2, and
  panel c shows transductive versus inductive thresholding for EG3.}
\caption{\textbf{Per-gap sensitivity heatmaps.}
  Cell values report the main effect size for each setting: relative
  \ap swing in (a), threshold-induced \tpr/\cdr range in percentage
  points in (b), and inductive-minus-transductive metric deltas in
  percentage points in (c).  Smaller text in
  (a) reports the window with the highest \ap.}
\label{fig:sc1-sc3-sensitivity}
\label{fig:snapshot-sensitivity}
\label{fig:threshold-instability}
\label{fig:transductive-thresholding}
\end{figure*}

Many \gids first divide events into snapshot windows and build one
graph per window.  Short windows yield sparse graphs that may split
related attack steps, while long windows yield dense graphs that can
hide short-lived attack signals in normal activity, so the snapshot
window is a model input, not an implementation detail.

Four of the five systems depend on such a window.  \Argus, \Euler, and
\Vgrnn construct fixed-duration graph snapshots.  \Pikachu's temporal
preprocessing uses a window to assign events to snapshots and align
training with the first attack snapshot.  \AnomalE is excluded as a
static-graph baseline.  For each windowed system we sweep four snapshot
window sizes per dataset: 30\,min--2\,h for LANL, 6\,min--150\,min for
OpTC, 10\,min--50\,min for CI-2017, and 75\,s--450\,s for Campus.  Among
the studied systems, \Euler and \Argus already report window-size
ablations in their original evaluations.  Our sweep asks what a
single-system ablation cannot answer, which is whether the sensitivity
is model-specific or systemic.

We measure sensitivity as the relative \ap swing across the tested
windows, computed as $(\text{best}-\text{worst})/\text{best}$.
Figure~\ref{fig:snapshot-sensitivity}(a) shows the effect is not limited
to one detector or dataset: 11 of 16 detector--dataset pairs exceed a
20\% relative \ap swing, with mean 38.3\%.  The largest absolute
changes are enough to change the conclusions of an evaluation:
\Argus on OpTC falls from 0.760 at its 6-minute default to 0.033 at a
150-minute window, \Pikachu on \cicids changes by 0.358 \ap, and
\Euler on OpTC changes by 0.259 \ap.

The best window is not consistent across detectors or datasets.  On
OpTC, the best window is 6\,min for \Argus and \Euler but 150\,min for
\Vgrnn and \Pikachu.  On Campus, it ranges from 75\,s for \Vgrnn to
450\,s for \Argus and \Pikachu.  No single snapshot window is uniformly
safe: a window that improves one system can degrade another, and a
window that works on one dataset may be a poor choice on another.  The
risk grows when a detector has multiple preprocessing and temporal
modules, because the same window choice can simultaneously alter graph
density, temporal alignment, and the inputs passed to downstream
encoders.

If a paper reports only one snapshot window, or omits it entirely,
readers cannot tell whether the reported accuracy comes from the
detector or from a favorable window.  This problem is especially
important for \ap, our primary threshold-free metric under rare
attacks.  \auc is less sensitive (mean relative swing 12.7\%), but it
also hides false-alert burden that matters in intrusion detection.

\paragraph{Recommendations.}
A complete evaluation documents the snapshot window and its selection
rule per dataset, treats the window as a tunable parameter,
characterizes sensitivity across sizes, and
never derives the window from the test set.  When possible,
evaluations should include snapshot-free or streaming baselines
to separate temporal discretization from detector architecture.

\begin{scbox}{EG2: Threshold-Selection Instability}{%
Detector scores become alerts only after a threshold is chosen.  When
that threshold is calibrated from a random validation sample, repeated
runs can move the operating point and substantially change event-level
and campaign-level detection.  The largest \tpr and CDR ranges in our
experiments reach 70.4 and 42.1 percentage points.}{2/5 systems}
\end{scbox}

Post-threshold metrics (\tpr, \fpr, precision, $F_1$, CDR) depend on a
detection threshold that converts anomaly scores into binary alerts, so
the threshold is part of the evaluation protocol: two runs with similar
score rankings can produce different alerts if they choose different
thresholds.

\Argus and \Euler calibrate this threshold from a random validation
sample: both sample a subset of training events, compute an
anomaly-score distribution on that subset, and use a percentile of that
distribution as the threshold.  Since the sample changes with the seed,
the threshold also changes, even when the detector, dataset, snapshot
window, and remaining pipeline settings are fixed.  We exclude
\Pikachu and \Vgrnn because they do not use this rule (\Pikachu has a
separate thresholding issue covered in EG3), and \AnomalE because it is
a static-graph detector without snapshot-level threshold calibration.

We run \Argus and \Euler on all four datasets at their default EG1
windows, with ten seeds per detector--dataset pair, and measure each
metric's range across seeds.
Figure~\ref{fig:threshold-instability}(b) shows the resulting spread.  The
instability is not limited to one detector or dataset.  \Euler on Campus
has a \tpr range of 70.4 percentage points, and \Euler on OpTC has a \tpr
range of 48.0 and a CDR range of 28.6 percentage points.  \Argus on LANL
has a \tpr range of 35.6 and a CDR range of 42.1 percentage points,
with up to 24 attack campaigns changing between detected and missed
across seeds.  \Argus on CI-2017 has a CDR range of 30.8
percentage points.

This instability matters because it affects the operating point a
deployment would use.  A threshold-free metric hides the problem
because it evaluates score rankings before a boundary is chosen, while
\tpr, \fpr, precision, $F_1$, and CDR depend directly on that boundary.
When the threshold is selected from a small random validation subset,
the same system produces different alerts and detects different attack
campaigns across runs, making the evaluation hard to reproduce or
compare against another system.

\paragraph{Recommendations.}
Threshold selection should be deterministic and reported as part of the
evaluation protocol.  Systems should choose the validation set once and
report which time period it covers rather than drawing a new random
subset per run.  Determinism does not guarantee that the chosen
threshold transfers.  A fixed validation period can be unlike the test
period it precedes: on \lanl, 12.5\% of validation edges are malicious
against 0.6\% of test edges, so a threshold well calibrated on
one is not automatically well calibrated on the other.  What
determinism buys is that the residual error is a property of a stated
protocol rather than an unreported seed, and that the operating
points in Table~\ref{tab:detection:main} can be reproduced exactly.
Papers should report the threshold-selection
rule, the validation data used for calibration, and the final threshold.  If random calibration is unavoidable, evaluations should report
variance across seeds for all post-threshold metrics,
including CDR.

\begin{scbox}{EG3: Transductive Thresholding}{%
A threshold chosen from evaluation data is not a deployable threshold.
When the threshold is fixed before the test period, the same detector can
lose campaign coverage entirely on some datasets and move to a very
different operating point.}{1/5 systems}
\end{scbox}

\Pikachu selects its anomaly threshold from the evaluation scores
themselves: the detector computes anomaly scores for post-training
edges and then chooses the threshold from those same labeled edges.
This is a transductive
protocol~\cite{pendlebury2019tesseract} because the threshold is fitted
to the data on which the detector is later evaluated, a choice
unavailable in deployment, where the threshold must be chosen before
the future test period is observed.  Selecting the operating point
from the test period is a lapse of basic machine-learning evaluation
hygiene rather than a failure specific to \gids.  The data-snooping
pitfall is already documented in security
evaluation~\cite{arp2022dos}.  We examine it here because \Pikachu is
the only one of the \numsystems studied systems that uses this
protocol and because its effect on reported results is large.

We compare this transductive protocol with an inductive alternative.
The inductive protocol splits the post-training snapshots
chronologically, chooses the threshold on the first half, and keeps
it fixed for the remaining test snapshots.

Figure~\ref{fig:transductive-thresholding}(c) shows that the protocol
change substantially shifts reported behavior.  On LANL,
\cdr drops from 59.6\% to 0.0\% and \tpr drops by 66.4 percentage
points.  On CI-2017, \cdr also drops to 0.0\% while \tpr and $F_1$
decrease by 83.6 and 75.3 percentage points.  The original threshold is
therefore not a post-processing detail: it can determine whether any
attack campaign is detected at all.

\begin{figure}[!t]
\centering
\includegraphics[width=\columnwidth]{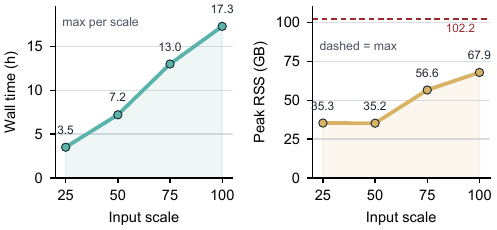}
\Description{Two-panel line chart showing the maximum EG5 wall-clock time
  and peak resident memory at 25, 50, 75, and 100 percent input scale.
  A dashed horizontal line marks the largest peak memory observed across
  all measurements.}
\caption{Runtime and memory scaling in EG5.  Points report the largest
value over detector--dataset pairs at each input scale.  The dashed
line marks the largest peak RSS observed across all measurements.}
\label{fig:sc5-scaling-cost}
\end{figure}

\begin{figure}[!t]
\centering
\includegraphics[width=0.9\columnwidth]{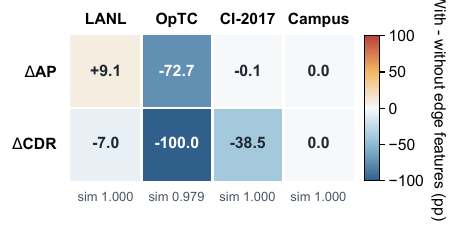}
\Description{Heatmap of with-feature minus without-feature AP and CDR
  deltas for Argus on the four datasets, with the attack to benign
  edge-feature cosine similarity printed once per dataset below the
  heatmap.}
\caption{EG4 edge-feature effect.  Cells report with-feature minus
without-feature \ap and \cdr for \Argus in percentage points.
Smaller text gives attack--benign edge-feature cosine.}
\label{fig:edge-feature-effect}
\end{figure}

The change is not always a simple decrease: on Campus,
\tpr and \fpr barely move under the inductive threshold while
precision drops by 54.5 percentage points, and on OpTC, \tpr and \cdr
remain unchanged while \fpr rises by 34.7 percentage points.  Transductive calibration tunes
the threshold on the test period while inductive calibration fixes it
beforehand, so the reported result reflects a more permissive
protocol rather than a deployable one.

\paragraph{Recommendations.}
Threshold calibration should be separated from test evaluation: a
paper should state where the threshold comes from, whether labels are
used, and whether it is fixed before the test period, with labeled
calibration data from a known time range before the test data.
Transductive thresholding should be reported as an oracle
setting, not as the main deployment result.

\begin{scbox}{EG4: Untested Edge-Feature Value}{%
Evaluations treat edge features as a source of signal without
isolating the feature path or testing whether it separates attack from
benign traffic.  Isolating it in \Argus changes results in
dataset-dependent ways and can remove campaign detection entirely,
while attack and benign feature means stay near-collinear on every
dataset.}{2/5 systems}
\end{scbox}

\Argus is the only system that provides an explicit edge-feature
switch, isolating these features without changing the detector
architecture.  We compare two versions of \Argus on each dataset, one
using the original edge features and one with the edge-feature path
disabled, with all other details fixed.  \AnomalE carries the mark as
well: its featurization is built on flow attributes that its released
evaluation never isolates, and it exposes no comparable switch, so the
same ablation cannot be run for it.  The remaining three systems
consume no edge features at all.

As shown in Figure~\ref{fig:edge-feature-effect}, edge features are not
a reliable source of improvement for \ap or \cdr.  On OpTC, enabling
edge features at the default 6-minute window reduces \ap from 0.760 to
0.033 and \cdr from 100.0\% to 0.0\%.  This comparison and the EG1 window
sweep both depart from the same baseline configuration (6-minute
window, feature path off), so the 0.760 endpoints coincide while the
two 0.033 endpoints are distinct measurements of different
single-factor changes.  On CI-2017, \ap remains near zero
but \cdr drops from 65.4\% to 26.9\%.  LANL shows the opposite metric
pattern: \ap improves from 0.014 to 0.105 while \cdr still drops from
71.9\% to 64.9\%.  Campus changes little in \ap and detects no
campaigns either way.  Adding edge features does not simply make the
detector stronger: it can change which attacks are found and in some
cases removes campaign-level coverage entirely.

\begin{figure}[!t]
\centering
\includegraphics[width=\columnwidth]{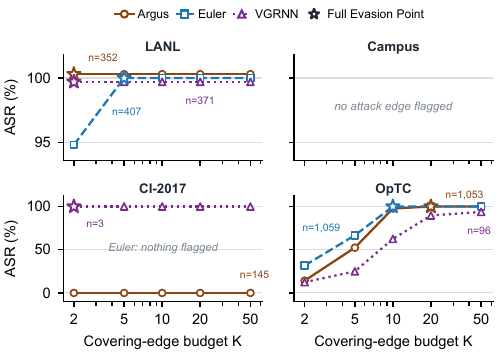}
\Description{Four-panel line chart showing attack success rate for Argus,
  Euler, and VGRNN on Campus, CI-2017, LANL, and OpTC as the
  covering-edge budget increases.  Stars mark the smallest budget that
  reaches full evasion, and each panel lists the per-detector flagged
  counts.  Panels or series with no line indicate that the detector
  flagged no attack edge.}
\caption{Adversarial covering-edge evasion in EG6.  ASR is computed
over the attack edges each detector flags, whose count $n$ is printed
beside its curve.  A star marks the smallest budget at which no
flagged edge survives.  A detector with no line flagged no attack
edge, so the attack has nothing to evade.  The LANL panel uses a
magnified axis, and its saturated curves carry a small vertical offset
so all three stay visible.}
\label{fig:sc6-perturbation}
\end{figure}

The mean attack and mean benign edge-feature vectors are nearly
collinear on every dataset (cosine 0.979 to 1.000, in the
smaller labels of Figure~\ref{fig:edge-feature-effect}).  Because these
features are non-negative, collinearity of means is weak evidence on
its own; the ablation above is what shows the feature path is not a
reliable signal.

\paragraph{Recommendations.}
Evaluations should not assume edge features are beneficial just because
they add traffic attributes.  Authors should report ablations with and
without edge features, and test whether attack and benign feature
distributions are separable before treating flow attributes as reliable
detection signal.  Feature overlap is not only a dataset artifact: an
adversary can force it.  In a replay attack, recorded benign flows are
reused to carry malicious interactions, so edge-feature values match
the benign distribution by construction rather than by accident.
Under such an attack the feature path carries no signal by design, and
the remaining evidence is structural: which hosts interact, and how
often.  Evaluations should therefore include replayed attack traffic
alongside standard attacks and verify that detection does not
depend on flow attributes alone.

\begin{scbox}{EG5: Scalability Limitations}{%
Accuracy-only evaluations hide the cost of running \gids at scale.
Even on benchmark workloads with up to 49.3M input events, our runs can
take more than 17 hours and reach 102.2\,GB peak memory, making such
costs hard to sustain on continuous enterprise traffic.}{5/5 systems}
\end{scbox}

Most \gids evaluations focus on accuracy and report little about
resource cost, but graph-based detectors must build large graph inputs,
extract features, train or apply models, and compute anomaly scores,
all of which grow with event volume.  A detector that is accurate on a
fixed benchmark may still be hard to use if its runtime or memory grows
quickly with the event stream.

We measure this cost by scaling detector-ready inputs.  For each
detector--dataset pair, we hold the detector's default snapshot window
fixed and vary only the input scale across 25\%, 50\%, 75\%, and 100\%
of the corresponding workload, recording wall-clock time and peak
resident memory. Figure~\ref{fig:sc5-scaling-cost} shows scalability is not a minor
detail.  At 100\% scale, the largest benchmark contains 49.3M input
events, the longest run takes 17.3 hours, and peak memory in the scale
experiment reaches 67.9\,GB.  Scaling is uneven: the largest
25\%--100\% runtime growth is 14.9$\times$ and the largest memory
growth is 4.5$\times$.  Across all measurements, including the EG1
snapshot-window experiments, peak memory reaches 102.2\,GB.

\begin{table}[!t]
\centering
\caption{Best-performing option in each \gids design dimension.  For
each dataset and each dimension, we average \ap over all configurations
that use the same option and report the option with the highest average
\ap.  Identity denotes a fixed Gaussian embedding keyed to the node
identity.  No single feature choice, encoder, temporal model, or
training target is best across all datasets.}
\label{tab:sc7-frameworked-ablation}
\scriptsize
\setlength{\tabcolsep}{3pt}
\renewcommand{\arraystretch}{1.10}
\begin{tabular}{@{}l l l l l@{}}
\toprule
\textbf{Dimension} & \textbf{Campus} & \textbf{CI-2017} & \textbf{LANL} & \textbf{OpTC} \\
\midrule
Featurization     & id.+struct.\ (.216)     & struct.\ (.553)        & struct.\ (.239)        & identity (.987)      \\
Graph encoder     & MPNN (.269)             & MPNN (.601)            & none (.254)            & GraphSAGE (.971)     \\
Temporal enc.     & GRU (.198)              & LSTM (.555)            & none (.210)            & GC-LSTM (.972)       \\
Objective         & reconstr.\ (.427)       & reconstr.\ (.782)      & reconstr.\ (.269)      & link pred.\ (.989)   \\
\bottomrule
\end{tabular}
\end{table}

\begin{figure}[!t]
\centering
\includegraphics[width=0.95\columnwidth]{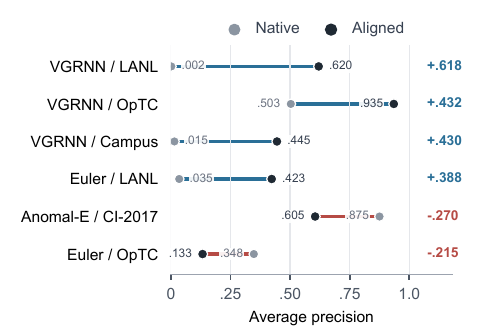}
\Description{Dumbbell chart comparing native and aligned average
  precision for the six detector and dataset pairs with the largest
  shifts under the shared preprocessing contract.}
\caption{Impact of preprocessing alignment in EG8 for the six
detector--dataset pairs with the largest \ap shifts.
Table~\ref{tab:app-sc8} lists all 18.  Native \ap uses each system's
own pipeline, aligned \ap the same detector under the
shared contract.}
\label{fig:sc8-preprocessing-impact}
\end{figure}

Reduced workloads can make a detector look more practical than it is
at full scale.  The largest runtime and memory growth occur on
different detector--dataset pairs, so scalability is not a single
system-wide constant, and a result on a reduced workload does not
reliably predict full-scale cost.  Peak memory also depends on
preprocessing choices outside the core model: EG5 holds the snapshot
window fixed to isolate input-scale effects, but EG1 shows that changing
the snapshot window can raise memory further.  The additional maximum
marker in Figure~\ref{fig:sc5-scaling-cost} reflects this gap: the largest
memory across all measurements is substantially higher than the largest
memory under the EG5 default-window setting.

Table~\ref{tab:sc-summary} marks a system for EG5 when its original
evaluation does not report resource cost and how that cost grows with
input scale, and all five systems carry the mark on those reporting
grounds.  Our measurements separate the reporting gap from the growth
itself, which is uneven.  The empirical runtime exponents in
Table~\ref{tab:app-sc5} stay at or below 0.6 for \Euler and 0.8
for \Vgrnn on every dataset, and \AnomalE is close to linear on the
two datasets where it runs, so these systems scale sub-linearly to
linearly in our sweep.  The one case above the 1.5 super-linear
threshold is \Pikachu on LANL, whose exponent of 2.0 corresponds to
the 14.9$\times$ runtime growth noted above.  Benign scaling in our
sweep does not clear the mark: without
reported cost, an evaluation cannot show that a detector remains
affordable at deployment scale.

\paragraph{Recommendations.}
\gids evaluations should report resource usage together with accuracy.
At a minimum, papers should report input scale, snapshot window,
wall-clock time, peak memory, and hardware configuration for each
evaluated detector.  When reduced workloads are used, evaluations
should also report how runtime and memory change with input scale
rather than treating scalability as an implementation detail.

\begin{scbox}{EG6: Unevaluated Adversarial Robustness}{%
Adversarial robustness is missing from \gids evaluation: the studied
systems are assessed on clean traffic only.  Instantiating a single
covering-edge attack shows what this omission hides.  Four of the 12
detector--dataset pairs flag no attack edge to begin with.  Of the eight
that do, six lose every flagged edge within 20 inserted edges and three
within only two.}{3/5 systems}
\end{scbox}

The gap here is broader than any single attack: the original
evaluations of the systems we study assess detection on clean traffic
only, so their deployability claims say nothing about an adversary who
adapts.  One attack cannot settle how robust these systems are, but it
can show what the missing evaluation hides.

We demonstrate this with the covering-edge evasion attack of Xu
\etal~\cite{xu2023cover}, which recent R+R work~\cite{wang2025rr} uses
on \gids as a state-of-the-art robustness test.  It applies when a
detector scores suspicious behavior as graph edges and an adversary can
add benign-looking covering edges during evaluation.  In \gids those are
network or host interactions inserted around an attack event to change
the graph evidence the detector sees.  From a flagged target edge, the
attack inserts covering edges until the target is stealthy while
the inserted edges stay undetected.  This is no random edge-insertion
test: it uses the detector's own graph-scoring path to pick
perturbations that reduce the evidence around the target.

For each detector--dataset pair we take every attack edge in the test
period as a candidate and record whether the detector flags it at that
pair's own threshold: the threshold of the released checkpoint where
the authors publish one (LANL, OpTC), and the threshold our
reproduction calibrates otherwise (Campus, CI-2017).  An edge already missed
cannot be evaded, so we run the attack only on the flagged ones, with
budgets $K \in \{2,5,10,20,50\}$.  Attack success rate (ASR) is the
fraction of \emph{flagged} targets made stealthy within $K$, so full
evasion, ASR 100\%, means no edge the detector caught survives.  A
detector that flags nothing has no ASR to report.

We instantiate it for \Argus, \Euler, and \Vgrnn, whose released
implementations expose graph-scoring paths suitable for
gradient-guided insertion.  Its absence for \Pikachu and \AnomalE is
not evidence of robustness: the protocol simply cannot be instantiated
against their released evaluation paths.

How much there is to evade varies sharply.  Each attack run uses the
detector's own preprocessing, default window, and calibrated threshold,
not the shared contract of Table~\ref{tab:detection:main}.  On Campus
none of the
three detectors flags an attack edge at its own threshold, and on
CI-2017 \Euler flags none, so on those four pairs there is
nothing to hide.  \Vgrnn on CI-2017 flags 3 of 119.  The eight
pairs with a flagged target are those an evasion budget can be
measured on.

Figure~\ref{fig:sc6-perturbation} shows that small budgets suffice on
most of them.  Two covering edges hide every flagged attack edge for
\Argus on LANL, across 352 targets, and for \Vgrnn on LANL across
371.  \Vgrnn on CI-2017 also falls at two edges, over its three
flagged targets.  \Euler on LANL reaches full evasion at $K=5$ over 407 targets,
\Euler on OpTC at $K=10$ over 1{,}059, and \Argus on OpTC at $K=20$
over 1{,}053.  Six of the eight measurable pairs therefore lose every
edge they caught within 20 inserted edges.  Because the attack runs
only on flagged edges, evasion means the covering edges changed a
decision rather than exploiting an existing miss: clean benchmark
accuracy is not sufficient evidence of robustness.

The two pairs that resist are also informative.  \Argus on CI-2017
converts none of its 145 flagged edges at any budget, and \Vgrnn on
OpTC rises from 12.5\% at $K=2$ to 93.8\% at $K=50$ without reaching
full evasion.  Adversarial robustness is therefore dataset and
system dependent and must be measured explicitly rather than inferred
from architecture or clean-data performance.

The capability this assumes divides in two.  The action is modest:
a covering edge is an innocuous-looking flow originating from a host
the adversary already controls, and two to twenty of them against millions
of benign events sit far below any practical rate limit.  The guidance
is not: choosing which edges to insert needs gradients through the
scoring path, so these budgets are a white-box upper bound on how cheap
evasion can be.  A weaker adversary needs a surrogate, and the
black-box evasion Wang \etal~\cite{wang2025rr} observe on \lanl
indicates that path is open.  Both variants assume the telemetry
faithfully records what crossed the network; an adversary who instead
alters the log is outside our threat model and is the concern of
tamper-evident auditing~\cite{custos,nitro}.

\paragraph{Recommendations.}
Adversarial robustness should be a standard axis of \gids evaluation
rather than an optional extra.  When a detector uses graph structure
during scoring, the evaluation should include covering-edge tests and
report the attack model, ASR across increasing budgets, and the
minimum budget for full evasion.  When an attack does not apply to a
design, that non-applicability should be stated explicitly and not
counted as evidence of robustness.

\begin{scbox}{EG7: Lack of Systematic Component Attribution}{%
GIDS evaluations often report end-to-end gains, but do not
systematically attribute those gains to individual pipeline components.
Local ablations can show that a proposed model degrades when modified,
but not whether the gain comes from graph reasoning, feature
construction, temporal modeling, the objective, or dataset-specific
signals.}{5/5 systems}
\end{scbox}

\begin{table}[!t]
\centering
\caption{Streaming-readiness replay results in EG9.}
\label{tab:sc9-streaming-readiness}
\small
\setlength{\tabcolsep}{5pt}
\renewcommand{\arraystretch}{1.4}
\begin{tabular}{lc}
\toprule
\textbf{Criterion} & \textbf{Result} \\
\midrule
Replay jobs & 54 \\
Detector--dataset pairs & 18 \\
Batch-only pairs & 14 \\
Near-real-time pairs & 4 \\
Real-time-ready pairs & 0 \\
Median buffering delay at 1$\times$ & 40 min \\
\bottomrule
\end{tabular}
\end{table}

Ablation is an attribution tool, not a checklist.  Existing \gids
papers may include local variants or sensitivity tests, but these are
scoped around one proposed architecture: they show whether a final
recipe is fragile to modification, not which part of the pipeline is
responsible for the result.  \Argus goes furthest, reporting a
component ablation over its own architecture, but even that stays
within one design.  A \gnn module may appear important because it
interacts with a particular feature representation, objective,
snapshot window, or dataset artifact, so the same end-to-end gain can
support several incompatible explanations. We therefore factorize \gids within \Sys into four interchangeable components,
featurization, graph encoder, temporal encoder, and training objective,
yielding \numconfigs configurations, each run on Campus, CI-2017,
LANL, and OpTC.  The goal is not to tune a new detector but to test whether
common \gids design choices have stable, attributable effects across
datasets.  Table~\ref{tab:sc7-frameworked-ablation} reports the best
marginal option in each dimension after averaging over the others.

\textbf{Graph structure provides dataset-dependent value.}
Removing the graph encoder, so that featurized inputs bypass message
passing entirely, is the best marginal choice on LANL and stays within
0.005 \ap of the best option on OpTC, while Campus and CI-2017 favor
MPNN.  Graph reasoning
can help, but does not reliably explain gains across datasets, and its
value must be demonstrated under controlled component variation.

\textbf{The training target changes the winner.}
Reconstruction is best on Campus, CI-2017, and LANL, and link
prediction is best on OpTC.  The same model architecture can look effective under one target
and weak under another, so without varying the target an evaluation
cannot tell whether the gain comes from the encoder or from what the
model is trained to predict.

\textbf{There is no universal GIDS recipe.}
Best choices change across datasets, and the winner is
an interaction among data representation, model structure, temporal
context, and objective.  Single-dataset ablations therefore do not
generalize: a component that helps on one benchmark may add little,
or even hurt, on another.

\paragraph{Recommendations.}
\gids evaluations should report systematic component attribution, not
just local ablations.  A minimum framework should vary the input
representation, graph encoder, temporal encoder, and training objective
under the same protocol, and include non-\gnn and non-temporal
baselines, since these controls are necessary to justify claims about
graph reasoning or temporal modeling.  Without this structure, reported
improvements may be real but their explanations remain unproven.

\begin{scbox}{EG8: No Common Preprocessing Contract}{%
GIDS papers often claim to evaluate systems on the same datasets, but
each system turns the raw logs into different model inputs.  These
pipelines change rows, features, graph semantics, labels, and splits, so
cross-system comparisons do not always measure the same benchmark.}{5/5 systems}
\end{scbox}

EG1 concerns one preprocessing parameter (the snapshot window) and EG4
isolates one feature-path switch in \Argus.  EG8 is a distinct
problem: the absence of a common preprocessing contract defining
how raw logs become model inputs.  Here preprocessing is no neutral
loading step: it defines the benchmark itself.

The same raw dataset can become different learning tasks: systems keep
different rows, extract different features, normalize differently,
collapse or preserve repeated edges, treat graphs as directed or
undirected, and attach labels to different units, so the pipeline
changes both the input and what the detector is asked to predict.

Prior cross-system comparisons show the problem.  When
\Argus~\cite{argus2020} reports \Euler as a baseline, their native
pipelines share no preprocessing contract, and even the systematic
re-eval\-u\-a\-tion of Wang \etal~\cite{wang2025rr} leaves each native
pipeline intact, so numbers from different studies need not measure the
same benchmark.

To measure the effect, we compare each system's native preprocessing
against an aligned contract that fixes a shared event table, graph
construction rule, label semantics, and split per dataset, and runs the
original detector on that common input.  This is not meant to tune the
detectors.  It asks whether the reported result is stable when the
preprocessing layer is held fixed.  Because EG1 shows the window moves
\ap on its own, both arms hold one window per dataset, common to every
system: 5\,min Campus, 50\,min CI-2017, 1\,h LANL and OpTC, so
preprocessing is the only variable.  A native score here is therefore
measured at that window rather than at the system's own default:
\Argus on OpTC scores .334 at 1\,h and .760 at the 6\,min default that
EG1 and EG4 sweep from.
Figure~\ref{fig:sc8-preprocessing-impact} shows the six largest \ap
shifts.

The \ap shifts are large enough to change the interpretation of
several systems.  \Vgrnn on LANL increases from .002 to .620 \ap under
the aligned contract.  \Vgrnn on Campus increases from .015 to .445,
and \Euler on LANL increases from .035 to .423.  The effect is not always
positive: \AnomalE on CI-2017 drops from .875 to .605, and \Euler on
OpTC drops from .348 to .133.  Preprocessing can therefore make a
detector look stronger or weaker before the model architecture is even
considered, and releasing only model code is insufficient: small
differences in row filtering, feature extraction, timestamp handling,
graph construction, or label assignment can move the evaluation target,
so the preprocessing path must be precise enough to reconstruct the
generated data byte-for-byte.

\paragraph{Recommendations.}
\gids evaluations should define the full raw-to-model preprocessing
contract.  At a minimum, papers should report the input schema, feature
extraction rules, graph construction rules, label unit, train/test
split, and snapshot policy for each dataset.  Authors should release
executable preprocessing scripts and, when feasible, checksums
or snapshots of the generated artifacts.  Cross-system comparisons
should report results under a shared aligned contract alongside
each system's native pipeline.

\begin{scbox}{EG9: Missing Real-Time Detection}{%
Existing \gids are evaluated as batch graph detectors rather than
streaming intrusion detectors.  They first buffer events into snapshots,
scenarios, or edge batches before scoring, so an attack cannot be
flagged when the malicious event arrives.}{5/5 systems}

\end{scbox}

EG5 asks whether a detector can finish processing a workload at scale,
whereas EG9 asks whether a newly arriving event can be scored immediately.  A
detector can have acceptable batch runtime and still be unusable for
real-time response if it must wait for a snapshot or scenario graph
before producing any alert.

We replay the archived executions from EG1, EG3, and EG5 under a
streaming arrival model.  The replay does not retrain the detectors.
It reuses each detector's runtime, input scale, batching unit,
and offline detection result, and simulates whether the detector
keeps up as events arrive at 1$\times$, 2$\times$, and 5$\times$ the
native ingest rate.  For each detector--dataset pair we record the
buffering delay before scoring begins, the time to the first
possible alert, the final processing lag, and whether the detector
keeps up with the replay stream.

Throughput alone does not imply real-time detection.
Table~\ref{tab:sc9-streaming-readiness} summarizes the replay outcome.
Across 54 replay
jobs covering 18 detector--dataset pairs, all jobs complete and most
systems keep up at the tested replay rates, yet none of the 18 pairs
is real-time ready.  Four short-window Campus pairs are only
near real-time (\Argus, \Euler, and \Vgrnn use 150-second windows,
while \Pikachu uses a 300-second window).  The remaining 14 are batch only.  The
median buffering delay at 1$\times$ replay is 40 minutes: alerts are
delayed by the need to form a graph input even when the detector can
process that input fast enough.

\Argus on LANL illustrates the pattern: it keeps up even at 5$\times$
replay speed but requires a 1-hour snapshot before scoring can begin.
The issue is not slow inference but that the unit of detection is
larger than the unit of arrival: real networks produce events
continuously, while current \gids score only after a larger graph
object has been formed.

\Euler comes closest to a streaming design among the studied systems:
it replicates the graph encoder across snapshot workers, with scalable
near-real-time detection as an explicit goal~\cite{king2022euler}.
Our replay bears out the throughput half of that design, since \Euler
keeps up at 5$\times$ the native ingest rate on all four datasets.  It
does not resolve the latency half: scoring still waits for a window to
close, so \Euler is near real-time only on the 150-second Campus
window and batch only elsewhere.

This makes current \gids closer to offline forensic analyzers than
real-time intrusion detectors.  They may still be useful for
retrospective analysis where delay is acceptable, but their
evaluations do not show timely response in a live network, where the
question is not only whether an attack is detected but when the alert
arrives.

\paragraph{Recommendations.}
\gids evaluations should report the detection unit and arrival unit
separately.  Papers should state whether the system scores edges,
nodes, snapshots, scenarios, or batches, and whether buffering is
required before scoring can begin.  When real-time detection is
claimed, authors should report event-to-alert latency under a streaming
arrival model, including snapshot window, buffer size, graph construction
delay, and per-event scoring time.  Batch-mode systems should be
described as offline or delayed detectors unless they can produce
alerts as events arrive.

\section{GIDS-Lite vs.\ Existing GIDS}\label{s:eval}

The analysis above motivates a simpler question: how much design
complexity is actually needed for \gids detection?  We answer with
\GidsLite, a deliberately simple baseline without graph or temporal
encoders, derived from the gaps of
\S\ref{s:shortcomings}~(\S\ref{s:gidslite}).  We then compare
\GidsLite with the \numsystems studied systems under the shared \Sys protocol
on detection (\S\ref{s:detection}) and computational cost
(\S\ref{s:cost}), isolate the effect of snapshot-window selection
(\S\ref{s:same-window}), and analyze adversarial covering-edge
robustness (\S\ref{s:gidslite-robustness}).

\subsection{GIDS-Lite Design}\label{s:gidslite}

\begin{table}[!t]
  \centering
  \small
  \setlength{\tabcolsep}{4pt}
  \renewcommand{\arraystretch}{1.15}
  \caption{\GidsLiteM candidates.  Mean \ap is averaged over the four
  validation splits, and the configuration with the highest mean is
  used unchanged on every dataset.}
  \label{tab:gidslite-impl}
  \begin{tabular}{@{}llrc@{}}
    \toprule
    \textbf{Endpoint state} & \textbf{Head} & \textbf{Dim.} &
    \textbf{Mean val.\ \ap} \\
    \midrule
    Identity emb.\ + struct.\ & Reconstruction & 296 & \textbf{0.809} \\
    Structural statistics     & Reconstruction &  40 & 0.791 \\
    Connectivity history      & Reconstruction &  40 & 0.763 \\
    Identity emb.             & Reconstruction & 264 & 0.632 \\
    Structural statistics     & Contrastive    &  40 & 0.351 \\
    Connectivity history      & Contrastive    &  40 & 0.349 \\
    Identity emb.\ + struct.\ & Contrastive    & 296 & 0.334 \\
    Identity emb.             & Contrastive    & 264 & 0.283 \\
    \bottomrule
  \end{tabular}
\end{table}

\paragraph{Purpose.}
\GidsLite is a minimal reference point for \gids evaluation: it keeps
the parts needed to predict from network events and drops the encoders
whose value should be demonstrated rather than assumed.  A strong
\GidsLite result does not show graph reasoning is never useful.  It
shows that a more complex \gids must provide clear gains over a simple
detector under the same protocol.

\paragraph{Design principles.}
\GidsLite rests on three principles.  It makes no evaluation-time
choices that move the operating point after seeing test data, so
thresholds are calibrated on a fixed chronological validation split.
Its model is simple enough to serve as a control, so a direct
predictor replaces the graph and temporal encoders.  It keeps the
same modular pipeline as the studied systems, so the shared \Sys
featurization and thresholding protocol holds.

\begin{figure}[!t]
\centering
\includegraphics[width=\columnwidth]{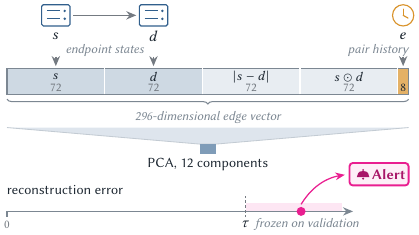}
\Description{GIDS-Lite architecture. Source and destination endpoint
states and eight pair-history scalars are concatenated with their
interaction terms into a 296-dimensional edge vector, scored by a
12-component PCA reconstruction head, and the frozen threshold turns
the anomaly score into an alert.}
\caption{\GidsLite architecture.  Endpoint states and pair-history
scalars form a 296-dimensional edge vector, which a 12-component PCA
reconstruction head scores.  The same configuration is used on all
\numdatasets datasets.}
\label{fig:gidslite-architecture}
\end{figure}

\paragraph{Model.}
\GidsLite contains no neural network, no message passing, and no
recurrence.  Its predictor is a PCA projection (12 components, fixed
seed) fitted to the edge vectors of the earliest part
of the trace, and each event is scored by its reconstruction error.
Shallow is meant literally: zero hidden layers, and no graph or
temporal encoder in the pipeline.
Figure~\ref{fig:gidslite-architecture} summarizes it.

One configuration is used on all \numdatasets
datasets.\footnote{Choosing a head per dataset would turn the control
into a tuned competitor: a per-dataset contrastive head reaches 0.79
\ap on \lanl, above the entry a single configuration earns in
Table~\ref{tab:detection:main}.  We report the single
validation-selected configuration precisely so that no per-dataset
choice enters the control, and accept the lower \lanl score that
follows.  The resulting split sizes are given in
Appendix~\ref{app:datasets}.}  We fix it
with a rule that never reads the test period: among the encoder-free
candidates of Table~\ref{tab:gidslite-impl}, which pair four endpoint
representations with a reconstruction or contrastive head, we take
the one with the highest mean \ap over the four validation splits.
That rule selects the 64-dimensional identity embedding concatenated
with eight structural statistics, scored by the reconstruction head, and we
use it unchanged on every dataset.  \GidsLite is thus not tuned
per dataset, which makes it a control, not a competitor:
each studied \gids keeps its published architecture, and
\GidsLite keeps one.

\paragraph{Features.}
Each event is represented by a numeric edge vector
$[\,s,\; d,\; |s-d|,\; s\odot d,\; e\,]$, z-scored on training data,
where $s$ and $d$ are the states of the source and destination
endpoints and $e$ collects eight edge-level scalars describing pair
history and recency (prior pair count, time since last seen, common
neighbors, in/out imbalance).  The endpoint state is a fixed
64-dimensional Gaussian embedding keyed to the node identity,
concatenated with eight structural statistics (degree, reciprocity,
neighbor statistics), giving a 296-dimensional edge vector on every
dataset.  \GidsLite does not
consume the traffic-attribute edge features that EG4 shows to be
collinear across attack and benign traffic.  Its signal comes from
connectivity and pair history, so it avoids those features.

\paragraph{Training and calibration.}
Training uses no attack labels.  Each head is fit on the
chronologically earliest snapshots of the trace, with a validation
fraction of 0.33 and a cap of 50{,}000 training edges, and the
detection threshold maximizes \mcc over a 201-point grid on the
chronological validation split, then stays frozen for testing.  Both
steps run under fixed random seeds, so training and calibration are
deterministic end to end.  The fitted model is correspondingly small:
a $12\times296$ projection plus a 296-dimensional mean, about 3{,}800
stored values in total.

\paragraph{Scope and granularity.}
\GidsLite performs no link prediction.  It assigns an anomaly score to
every observed event, so decisions are per edge rather than per node
or per graph.  Because its features derive from snapshot structure,
\GidsLite buffers the current snapshot before scoring and therefore
shares the EG9 constraint with the studied systems.  We claim
per-event scoring, not real-time detection.  Throughout, graph-based
refers to the input representation rather than to a graph model, and
\GidsLite is the encoder-free control within that setting.
Appendix~\ref{app:datasets} states, for each dataset, what
constitutes an edge and how labels are assigned.

\paragraph{Response to the evaluation gaps.}
\GidsLite is a control baseline, not a universal replacement for \gids
designs.  Its purpose is to test whether the additional complexity in
existing systems provides measurable value once the evaluation protocol
is held fixed.  Table~\ref{tab:sc-summary} summarizes the resulting
coverage: \GidsLite closes six of the \numgaps gaps, EG2, EG3, and EG8
through the shared protocol, EG4 by avoiding the feature path, EG5
through the cost reported in \S\ref{s:cost}, and EG7 through component
attribution.  Two remain open.  Window sensitivity (EG1,
\S\ref{s:same-window}) persists for \GidsLite, and snapshot buffering
leaves EG9 unresolved.  The
covering-edge attack of EG6 cannot be instantiated against a detector
without a graph-scoring path (\S\ref{s:gidslite-robustness}).  This is
about attack applicability, not general robustness.

\begin{table}[!t]
\centering
\caption{Evaluation-gap status across the studied systems and
\GidsLiteM. \xmark{} marks a gap that remains open for the pipeline,
\cmark{} one that its protocol or design closes, and a dash a gap
whose test cannot be instantiated for that design;
\S\ref{s:shortcomings} states the criterion applied for each gap.}
\label{tab:sc-summary}
\scriptsize
\setlength{\tabcolsep}{3pt}
\begin{tabular}{@{}lccccccccc@{}}
\toprule
\textbf{System} & \textbf{EG1} & \textbf{EG2} & \textbf{EG3} & \textbf{EG4} & \textbf{EG5} & \textbf{EG6} & \textbf{EG7} & \textbf{EG8} & \textbf{EG9} \\
\midrule
\ArgusM    & \xmark & \xmark & \cmark & \xmark & \xmark & \xmark & \xmark & \xmark & \xmark \\
\EulerM    & \xmark & \xmark & \cmark & \cmark & \xmark & \xmark & \xmark & \xmark & \xmark \\
\PikachuM  & \xmark & \cmark & \xmark & \cmark & \xmark & --     & \xmark & \xmark & \xmark \\
\VgrnnM    & \xmark & \cmark & \cmark & \cmark & \xmark & \xmark & \xmark & \xmark & \xmark \\
\AnomalEM  & \cmark & \cmark & \cmark & \xmark & \xmark & --     & \xmark & \xmark & \xmark \\
\midrule
\GidsLiteM & \xmark & \cmark & \cmark & \cmark & \cmark & -- & \cmark & \cmark & \xmark \\
\bottomrule
\end{tabular}
\end{table}

\subsection{Detection Performance}\label{s:detection}

Using the protocol of \S\ref{s:framework}, we compare \GidsLite with
the \numsystems studied systems under the same preprocessing,
chronological-split and threshold-calibration rules, and reporting
pipeline.
Table~\ref{tab:detection:main} reports the end-to-end comparison on all
four datasets.

{\renewcommand{\arraystretch}{1.14}%
\begin{table*}[!t]
  \centering
  \scriptsize
  \setlength{\tabcolsep}{2.8pt}
  \caption{Detection results under the shared \Sys protocol.  \cdr
    counts the campaigns in each pipeline's test period, and the
    denominators of \AnomalEM and \GidsLiteM differ from the
    shared-split campaign sets of the studied systems.
    \AnomalEM is not applicable to \lanl and \optc, whose
    authentication and process events carry none of the
    flow attributes it consumes.}
  \label{tab:detection:main}
  \resizebox{\textwidth}{!}{%
  \begin{tabular}{l cccc cccc cccc cccc}
    \toprule
    & \multicolumn{4}{c}{\textbf{Campus}}
    & \multicolumn{4}{c}{\textbf{\lanl}}
    & \multicolumn{4}{c}{\textbf{\cicids}}
    & \multicolumn{4}{c}{\textbf{\optc}} \\
    \cmidrule(lr){2-5} \cmidrule(lr){6-9} \cmidrule(lr){10-13} \cmidrule(lr){14-17}
    \textbf{System}
      & \textbf{\ap} & \textbf{\cdr} & \textbf{Prec.} & \textbf{\fpr}
      & \textbf{\ap} & \textbf{\cdr} & \textbf{Prec.} & \textbf{\fpr}
      & \textbf{\ap} & \textbf{\cdr} & \textbf{Prec.} & \textbf{\fpr}
      & \textbf{\ap} & \textbf{\cdr} & \textbf{Prec.} & \textbf{\fpr} \\
    \midrule

    \ArgusM
      & 0.017 & 0.000 & 0.000 & 0.001
      & 0.104 & 0.105 & 0.031 & 0.014
      & $<.001$ & 0.000 & 0.000 & 0.209
      & 0.521 & 1.000 & 0.439 & 0.009 \\

    \EulerM
      & 0.004 & 1.000 & 0.007 & 1.000
      & 0.423 & 0.649 & 0.498 & 0.024
      & $<.001$ & 0.000 & 0.000 & 0.042
      & 0.133 & 0.714 & 0.149 & 0.048 \\

    \PikachuM
      & 0.915 & 1.000 & 0.616 & 0.066
      & 0.059 & 0.702 & 0.071 & 0.404
      & 0.808 & 0.654 & 0.683 & 0.166
      & 0.987 & 1.000 & 0.994 & 0.006 \\

    \VgrnnM
      & 0.445 & 0.000 & 0.000 & $<.001$
      & 0.620 & 0.912 & 0.122 & 0.366
      & 0.280 & 0.000 & 0.000 & $<.001$
      & 0.935 & 1.000 & 0.742 & 0.231 \\

    \AnomalEM
      & 0.048 & 1.000 & 0.048 & 1.000
      & -- & -- & -- & --
      & 0.605 & 1.000 & 0.658 & 0.192
      & -- & -- & -- & -- \\

    \midrule

    \rowcolor{gidslitecolor!8}
    \GidsLiteM
      & 0.322 & 1.000 & 0.178 & 0.056
      & 0.510 & 0.529 & 0.245 & 0.013
      & 0.980 & 0.714 & 0.976 & 0.009
      & 0.993 & 1.000 & 0.986 & 0.587 \\

    \bottomrule
  \end{tabular}}
\end{table*}
}

By \ap, one fixed encoder-free configuration is the strongest system
on \cicids and \optc, where \GidsLite reaches 0.980 and 0.993 without
a graph encoder or a temporal encoder.  The other two datasets go the
other way: \Pikachu reaches 0.915 \ap on Campus against 0.322 for
\GidsLite, and \Vgrnn reaches 0.620 on \lanl against 0.510.  The
detection benefit of added complexity is therefore dataset dependent,
and the split is not the one the simplicity argument would predict:
complexity pays on the two datasets with the sparsest attack signal
and is redundant on the two where a linear reconstruction of
connectivity features already separates attack from benign.  Where it
does pay, the gain must still be weighed against the runtime cost
quantified in \S\ref{s:cost}.

Threshold-free and post-threshold results do not always move together,
and no system is uniformly good on both.  \Vgrnn has non-trivial \ap on
Campus and \cicids but zero \cdr and zero precision after thresholding
on both.  \GidsLite shows the same gap from the other side: its
thresholded operating point is attractive only on \cicids (97.6\%
precision at a 0.9\% \fpr).  On Campus it exposes its test period's single campaign at
17.8\% precision, on \lanl its precision is 24.5\% where \Euler
reaches 49.8\%, and on \optc it flags 59\% of the benign events, a
rate \Pikachu avoids while scoring almost as well.  Ranking quality
alone does not identify the system an operator would
deploy.

\begin{figure}[!t]
\centering
\includegraphics[width=0.88\columnwidth]{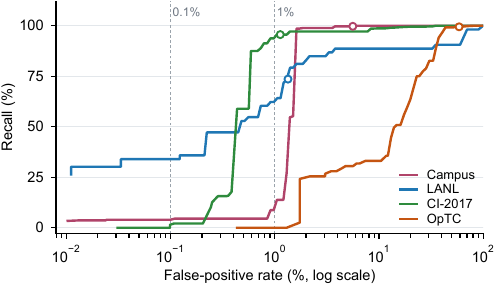}
\Description{Recall versus false-positive rate for GIDS-Lite on the four
  datasets, with the false-positive rate on a logarithmic axis and the
  reported detection operating point marked on each curve.}
\caption{\GidsLite recall against the alert budget.  Circles mark the
  operating point reported in Table~\ref{tab:detection:main}, and dashed
  lines mark the 0.1\% and 1\% \fpr budgets.  No dataset retains its
  reported recall once the budget tightens to 0.1\%.}
\label{fig:gidslite-recall-fpr}
\end{figure}

Reported detection rates should also be read against a realistic
alert budget, since a high false-positive rate can make an otherwise
accurate detector unusable in practice~\cite{sommer2010outside,
arp2022dos, nodoze2019}.  We therefore sweep the \GidsLite operating point and
report recall at analyst-relevant false-positive rates
(Figure~\ref{fig:gidslite-recall-fpr}).  These curves reflect the
single configuration of \S\ref{s:gidslite}, which is fixed on
validation data and never tuned per dataset.  No dataset survives a
strict budget.  At a 1\% \fpr, recall is 94\% on \cicids but 62\% on \lanl,
9\% on Campus, and zero on \optc.  At 0.1\% it is 34\% on \lanl and
under 4\% everywhere else.  The reported operating points in
Table~\ref{tab:detection:main} rest on alert budgets an
analyst would question: roughly 1\% on \cicids and \lanl, 5\% on
Campus, and tens of percent on \optc.  \optc also has a
97.6\% attack base rate in its test period, which is what lifts every
system's \ap there and is the reason we read its column as the least
informative of the four.  These trade-offs, not single-threshold
summaries, are what a deployment decision depends on.

Under a shared protocol, the return on added design complexity is
inconsistent rather than absent.  A single encoder-free configuration
leads on two of four datasets, and on the other two the systems that
win do so with different architectures, \Pikachu on Campus and \Vgrnn
on \lanl, so no one design is vindicated either.  The post-threshold
columns add a second caveat: strong ranking must still be checked
against campaign exposure and alert burden.

\subsection{Runtime and Memory}\label{s:cost}

We record end-to-end wall-clock time and peak resident memory for
every detector under the shared contract.  \GidsLite takes 107.2 seconds
and 11.0\,GB while \Pikachu takes 17.1 hours and several systems exceed
27\,GB.  Figure~\ref{fig:gidslite-cost-pareto} places every system on the
same time--memory cost plane.  As shown in \S\ref{s:shortcomings}, other valid
preprocessing choices can raise observed peak memory to 102.2\,GB.

The cost ranking does not match the detection ranking.  \Pikachu has
the highest wall-clock time, but \GidsLite still achieves stronger \ap
on multiple datasets, and \Euler and \Vgrnn incur substantially higher
memory without a consistent detection advantage.

\subsection{Snapshot-Window Sensitivity}\label{s:same-window}

EG1 shows that snapshot-window choice can substantially change the
reported performance of existing \gids.  We test whether \GidsLite
escapes it by holding the configuration fixed and varying only the
snapshot-window size over the EG1 window grid, with the
chronological split and validation-calibrated threshold of
\S\ref{s:detection}.

Table~\ref{tab:app-gidslite-window} reports the variation
across window sizes.  The largest \GidsLite \ap swing is 78.1\% on
\lanl and the smallest 0.8\% on \cicids, a range comparable to the
studied systems in EG1, and on \lanl campaign exposure moves as well,
by 23.5 percentage points of \cdr.  Removing the
encoders therefore does not remove the sensitivity: EG1 is a property
of turning a continuous stream into graphs, not of the architecture
that consumes them, and \GidsLite inherits it in full.  The two
datasets where \GidsLite ranks first are also the two where the window
barely matters, which is a reason to report the setting rather than a
reason to trust the ranking.

\subsection{Attack Surface}\label{s:gidslite-robustness}

EG6 shows that the cost of added architectural complexity is not
limited to computation: it can also include an added attack
surface.  Recent graph attacks~\cite{xu2022blindfolded, sharma2023tdap,
xu2023cover, zhao2025teani} exploit exactly such interfaces.  What they
share is a graph-scoring path that can be influenced by changing graph
structure, injecting graph elements, or probing graph-based model
responses.

In our codebase, the released implementations of \Argus, \Euler, and
\Vgrnn expose graph-scoring paths, so the EG6 covering-edge
attack~\cite{xu2023cover} applies directly to them.  \GidsLite scores
each edge from handcrafted snapshot features and a shallow predictor,
with no multi-hop aggregation at inference, so the EG6 attack cannot
be directly instantiated against it.  The narrow conclusion is not
that \GidsLite is adversarially robust in general, but that its
simpler scoring path avoids this graph-propagation attack
surface.

We therefore test the attack its design does admit.  Because \GidsLite scores each edge from
a standardized feature vector, an adversary who can shape those
features faces a minimum-perturbation problem, which we solve exactly
in the white-box setting: for every detected attack edge we compute the
smallest perturbation, in standardized feature units, that pushes the
edge below the frozen threshold.  Because this attacker is exact
rather than gradient-approximate, its budgets are an upper bound on
what \GidsLite withstands.  Starting from the operating points of
Table~\ref{tab:detection:main}, the median minimum perturbation is
12.4 on Campus, 3.3 on \lanl, 1.8 on \cicids, and 1.9 on \optc, and
the datasets where \GidsLite scores best are the ones an attacker
breaks most cheaply: a budget of two standardized units removes 78
percentage points of recall on \cicids and 60 on \optc, while Campus
loses one point.  Restricting the attacker to the pair-history
scalars, with the endpoint identity block held fixed, changes the
picture again: the median cost rises to 7.7 on \cicids and 4.7 on
\optc, and on Campus 98\% of the detected attack edges become
unevadable at any budget.  The attack therefore depends on being able
to move the endpoint representation, not the pair-history scalars
alone.  \GidsLite is not adversarially robust, and it
trades one attack surface for another.

More general evasion strategies~\cite{zhang2022advnids,
sharon2022tantra, shoaib2026catch}, from traffic-feature manipulation
to timing reshaping and semantics-preserving rewriting, remain
relevant for encoder-light detectors and are outside our scope.  The
narrower claim is about attack applicability: added graph and temporal
components introduce concrete attack surfaces that \GidsLite does not
expose, so architectural complexity should be judged not only by
clean-data accuracy and computational cost, but also by the attack
surface it introduces.

\section{Discussion \& Future Work}\label{s:discussion}

The main lesson of this study is not that simpler models always win.
Under a shared evaluation protocol, added graph and temporal complexity
must be justified by clear net benefit, assessed not only through
clean-data detection quality but also through computational cost,
stability, and attack applicability.

\paragraph{Relation to Wang \etal~\cite{wang2025rr} and Bilot
\etal~\cite{bilot2025simpler}.}
Our methodology adapts the staged PIDS decomposition of Bilot \etal to
network flows, extending it to per-component attribution across
systems so that symptoms Wang \etal could only observe end to end are
traced back to individual stages (EG7).  Our scope overlaps their
black-box re-evaluation;
Table~\ref{tab:prior-findings} (Appendix~\ref{app:prior-findings})
summarizes axis by axis which findings are new and which are not.
The two data models also differ in ways that shape what can
transfer: PIDS graphs inherit structure from kernel-event
causality~\cite{inam2022sok} and use symbolic features such as file
paths~\cite{flash2024}, whereas \gids must impose a snapshot
window on a continuous flow stream, rely on statistical edge
aggregates, and share no canonical preprocessing across papers.
\emph{Novel:} the cross-system snapshot-window sensitivity of EG1
(mean 38.3\% relative \ap swing: PIDS evaluations batch at fixed
durations they never vary, and Wang \etal flag the window without
quantifying it across systems), and the
contract-aligned preprocessing attribution of EG8 (\Euler on \lanl
moves from .035 to .423, where prior re-runs were per-paper and ad
hoc).  \emph{Inverts:} flow attributes do not carry the separating
signal that PIDS symbolic features do, and enabling \Argus's
edge-feature path on \optc drops \ap from 0.760 to 0.033 and \cdr from
100.0\% to 0.0\% (EG4).  Two covering edges fully evade the graph-scoring path on
\lanl, and six of the eight pairs with a flagged target fall within
twenty inserted edges, because the \gids graph is built from
attacker-observable traffic (EG6), which also explains the black-box
evasion Wang \etal observe on \lanl.  No single recipe wins across
datasets, against the one-recipe conclusion for PIDS (EG7).
\emph{Confirms:} published \gids
results are fragile when re-run, as Wang \etal report, and a
simple baseline is competitive once the evaluation is matched, as
Bilot \etal found for PIDS.  The threshold- and cost-hygiene concerns
of EG2, EG3, and EG5 are shared with both studies and sharpened
into per-pair measurements.  \emph{Adapts:} the deployment axis Bilot
\etal address for PIDS with VELOX~\cite{bilot2025simpler} becomes a
replay measurement for \gids, with none of 18 detector--dataset pairs
real-time ready (EG9).

\paragraph{Added Complexity Needs Clear Benefits.}
\S\ref{s:detection}--\S\ref{s:gidslite-robustness} show that
architectural complexity in current \gids is not free under our
protocol: added graph and temporal
components do not consistently improve clean detection, but they
increase runtime and memory cost, make the operating point sensitive
to threshold calibration, and may expose additional attack surfaces.
That does not make simple designs always preferable.  It makes
complexity a claim to be earned empirically rather than assumed.

\paragraph{Current Benchmarks Still Mix Key Factors.}
Each dataset bundles network setting, logging source, feature
visibility, attack type, and temporal granularity, so when a system
improves on one dataset but not another, the change may come from the
model, the features, the attack behavior, or the representation.
Future benchmarks should vary these factors deliberately while holding
the rest fixed.  Without it, claims about when graph structure helps
stay unresolved.

\begin{figure}[!t]
\centering
\includegraphics[width=0.73\columnwidth]{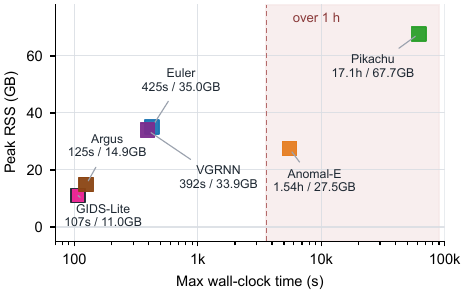}
\Description{Scatter plot of maximum wall-clock time versus peak resident
  memory for GIDS-Lite and the studied GIDS. Runs taking more than one
  hour are shaded.}
\caption{Computational cost under the EG8 shared preprocessing
contract.  Each point is the detector's largest wall-clock time and peak
resident memory across the \numdatasets datasets on the common input
(two for \AnomalEM); the \GidsLiteM point is its \cicids run in the EG7
grid.  Table~\ref{tab:app-sc5} instead prices each system on its
native pipeline.}
\label{fig:gidslite-cost-pareto}
\end{figure}

\paragraph{Interpretability Still Matters in Practice.}
Detection quality is only part of operational usefulness. Analysts
must also trace alerts to concrete evidence, which gets harder as
scoring spreads across graph and temporal components.  We do not
evaluate explanation quality, so future \gids evaluations should
include alert traceability alongside detection metrics.

\paragraph{Robustness Must Be Evaluated Explicitly.}
The adversarial result in \S\ref{s:gidslite-robustness} adds another
dimension.  Under the current protocol, several existing \gids expose
graph-scoring paths that a covering-edge attack can exploit, while
\GidsLite does not rely on that path.  The conclusion is not that
simpler detectors are universally robust, but that robustness cannot be
assumed as a by-product of model complexity.  Attack success and attack
applicability should be reported together.  Future work should study
detector-agnostic attack models, defenses for both graph-based and
encoder-light detectors, and robustness metrics that reflect the
NIDS threat model.

\paragraph{Shared Evaluation Reduces Repeated Work.}
Progress in \gids depends on evaluation infrastructure as much as on
models.  When papers use incompatible splits, preprocessing pipelines,
or reporting rules, the cost of rigorous comparison is paid many times
over without producing cumulative evidence.  Shared preprocessing
artifacts, canonical splits, common metric suites, and reusable
harnesses would make results comparable.  That infrastructure is part of
the research agenda, not auxiliary to it.

\section{Conclusion}\label{s:conclusion}

We presented \Sys, an evaluation framework that attributes \gids
performance to pipeline stages rather than whole systems and measures
\numgaps evaluation gaps current studies leave unexamined.  Under a matched protocol, added
graph and temporal complexity does not reliably buy detection quality,
yet it costs runtime and calibration stability and adds attack
surface, and
our encoder-free control, \GidsLitePlain, competes on half the
datasets at a fraction of the runtime.  Complexity in \gids should be
introduced only when it earns its cost under a shared protocol.

\begin{acks}
We sincerely thank our shepherd and the anonymous reviewers for their
insightful feedback on this work.  This material is based upon work
supported by the National Science Foundation (NSF) under grant numbers
\#2339483, \#2530655, and \#2622986.
\end{acks}

\balance
\bibliographystyle{abbrvnat}
{\footnotesize\sloppy\setlength{\emergencystretch}{1.5em}\setlength{\bibsep}{0.6pt}
\bibliography{bib-files/gids}}

\clearpage
\nobalance
\let\balance\relax
\appendix
\section{Open Science}
\label{app:open-science}

The paper artifact is available at
\url{https://github.com/DART-Laboratory/GIDS-Eval}.
It contains the \Sys framework, the rebuilt detector implementations,
the sweep configurations, the per-job results behind every reported
number, and setup instructions.  All \numdatasets datasets are publicly
available and are fetched from their original hosts by the setup
instructions.  Appendix~\ref{app:datasets} reports their statistics.

\section{Ethical Considerations}

This work involves no human subjects, live network experiments, or
production systems.  All experiments use public datasets, released
implementations, and offline traces, and the adversarial experiments
target no real credentials, live systems, or third parties.

\section{Generative AI Usage}

Claude (Anthropic) and GitHub Copilot assisted with editing and
coding.  All output was reviewed and revised.  The study design,
experiments, analysis, and conclusions are the authors' own.

\section{Reporting Notes}
\label{app:notes}

This appendix states the reporting conventions for
Appendix~\ref{app:per-sc}, which reports the full detector--dataset
matrix under the shared protocol of \S\ref{s:framework}.  Dashes mark non-applicable
settings, a default outside the swept grid, or an outcome not reached
within the tested range, and the surrounding text distinguishes them.
Wall-clock times use seconds below 60\,s, minutes below 3{,}600\,s,
and hours otherwise, and snapshot windows keep the unit of their sweep
grid.  Tables abbreviate them as s, m, and h.  Memory is peak resident
set size in GB.  All values derive from the artifact release that
produces
Table~\ref{tab:detection:main}, Figure~\ref{fig:gidslite-cost-pareto},
and Table~\ref{tab:app-gidslite-window}.  The artifact covers every
measurement, including the rebuilt systems, sweep configurations,
and analysis code.  Datasets are
fetched from their original hosts, and wall-clock and memory figures
are tied to the hardware of \S\ref{s:framework}.

\section{Dataset Details}
\label{app:datasets}

Table~\ref{tab:datasets} reports the per-dataset statistics referenced
in \S\ref{s:framework}.  The four datasets are chosen to vary across
deployment environment, telemetry type, scale, and attack-labeling
mechanism.  Campus contributes three controlled APT-style campaigns
(Oilrig, SandWorm, and WizardSpider).  CI-2017~\cite{liu2022error} is
the corrected-label variant of the original CIC-IDS-2017
dataset~\cite{cicids}.

Under the shared protocol, the unit of detection is defined per
dataset as follows.  On \lanl, an edge is a source-to-destination
authentication event, with no restriction to a particular
authentication protocol, and it is malicious if it matches the
red-team ground truth.  On \optc, an edge is a source-to-destination host
event, malicious if its time and endpoint pair match a red-team
event.  On CI-2017, an edge is a flow between a source and a
destination IP address, labeled directly by the corrected dataset
labels.  On Campus, an edge is a source-to-destination event from the
released R+R trace~\cite{wang2025rr}, malicious if it belongs to one
of the injected campaigns.

\begingroup\sloppy
Under the \GidsLite split rule of \S\ref{s:gidslite} (chronological,
validation fraction 0.33, 50{,}000-edge training cap), the resulting
train, validation, and test sizes are as follows.  Campus has
29{,}274 edges for training, 14{,}107${+}$384 benign${+}$attack for
validation, and 29{,}065${+}$355 for test.  \lanl has 506,
4{,}563${+}$649, and 8{,}971${+}$53.  CI-2017 has 50{,}000,
448{,}257${+}$184{,}482, and 762{,}531${+}$332{,}928.  \optc has
50{,}000, 12{,}597${+}$3{,}494, and 230${+}$9{,}389.\par\endgroup

\begin{table}[H]
  \centering
  \footnotesize
  \setlength{\tabcolsep}{4pt}
  \renewcommand{\arraystretch}{1.12}
  \caption{Datasets used in our evaluation.  Hosts/IPs counts unique
  endpoints by native identifier (host names for
  Campus, \lanl, and \optc, and IP addresses for CI-2017).}
  \label{tab:datasets}
  \begin{tabular}{@{}l r r r l@{}}
    \toprule
    \textbf{Dataset} & \textbf{Hosts/IPs} & \textbf{Events} &
      \textbf{Days} & \textbf{Attack labels} \\
    \midrule
    Campus~\cite{wang2025rr}     & 18.4M & 10.3B  & 101 & APT campaigns    \\
    \lanl~\cite{lanl}            & 17.7K & 1.18B  &  58 & Red team         \\
    \optc~\cite{optc}            &   814 & 10.1M  &   8 & Red team         \\
    CI-2017~\cite{liu2022error}  & 19.1K &  2.10M &   5 & Scripted attacks \\
    \bottomrule
  \end{tabular}
\end{table}

\section{Per-Gap Results}
\label{app:per-sc}

\subsection{EG1: Snapshot-Window Sensitivity}
\label{app:sc1}

Table~\ref{tab:app-sc1} extends Figure~\ref{fig:snapshot-sensitivity}(a)
with the default window for every pair next to the best one, the
corresponding absolute \ap values, and an \auc-swing column that the
figure does not display.  The default window is also the best window in only four
of the sixteen pairs, which is what motivates reporting both columns.
For \Pikachu on \cicids the native one-hour default lies outside the
10--50-minute grid swept on that dataset, so its default-window \ap is
dashed and its swing spans the tested windows.

\begin{table}[H]
  \centering
  \footnotesize
  \setlength{\tabcolsep}{2.5pt}
  \renewcommand{\arraystretch}{1.05}
  \caption{EG1 window sensitivity.  Swing is the relative gap between
  the best and worst tested window for the same detector--dataset
  pair, computed as $(\max - \min)/\max$.}
  \label{tab:app-sc1}
  \begin{tabular}{@{}l l c c c c r r@{}}
    \toprule
    \textbf{System} & \textbf{Dataset} & \textbf{Default} & \textbf{Best}
    & \textbf{Def. \ap} & \textbf{Best \ap}
    & \textbf{$\Delta$\ap} & \textbf{$\Delta$\auc} \\
    \midrule
    \multirow{4}{*}{\ArgusM}
      & Campus  & 150\,s & 450\,s & 0.001 & 0.001 &  0.6\% & 40.4\% \\
      & \lanl   & 1\,h    & 2\,h    & 0.105 & 0.177 & 89.5\% &  0.9\% \\
      & \cicids & 15\,m   & 10\,m   & 0.002 & 0.002 & 78.0\% & 41.9\% \\
      & \optc   & 6\,m    & 6\,m    & 0.760 & 0.760 & 95.7\% &  7.6\% \\
    \midrule
    \multirow{4}{*}{\EulerM}
      & Campus  & 150\,s & 150\,s & 0.001 & 0.001 &  0.7\% & 33.8\% \\
      & \lanl   & 30\,m  & 1\,h    & 0.029 & 0.035 & 21.0\% &  0.9\% \\
      & \cicids & 50\,m  & 10\,m   & 0.001 & 0.002 & 53.2\% & 27.9\% \\
      & \optc   & 150\,m & 6\,m    & 0.245 & 0.504 & 51.3\% &  1.8\% \\
    \midrule
    \multirow{4}{*}{\PikachuM}
      & Campus  & 5\,m   & 450\,s  & 0.896 & 0.922 &  4.0\% &  0.1\% \\
      & \lanl   & 1\,h   & 1\,h    & 0.091 & 0.091 & 30.7\% & 11.7\% \\
      & \cicids & 1\,h   & 50\,m   & --    & 0.843 & 42.4\% & 12.9\% \\
      & \optc   & 1\,h   & 150\,m  & 0.988 & 0.998 &  1.8\% &  0.9\% \\
    \midrule
    \multirow{4}{*}{\VgrnnM}
      & Campus  & 150\,s & 75\,s   & 0.021 & 0.023 & 33.9\% &  8.8\% \\
      & \lanl   & 90\,m  & 90\,m   & 0.003 & 0.003 & 72.4\% &  0.9\% \\
      & \cicids & 10\,m  & 30\,m   & 0.297 & 0.304 &  3.3\% & 10.6\% \\
      & \optc   & 30\,m  & 150\,m  & 0.512 & 0.517 & 34.0\% &  2.4\% \\
    \bottomrule
  \end{tabular}
\end{table}

\begin{table*}[!t]
  \centering
  \footnotesize
  \setlength{\tabcolsep}{5pt}
  \renewcommand{\arraystretch}{1.05}
  \caption{EG5 wall-clock time and peak resident set size at four
  input scales.  \emph{Exp.} is the growth exponent of wall-clock time
  between the 25\% and 100\% input scales, where 1.0 means runtime
  grows in proportion to input, and memory is in gigabytes.
  These runs use each system's \emph{native}
  pipeline, whereas Figure~\ref{fig:gidslite-cost-pareto} prices the
  shared-contract runs.}
  \label{tab:app-sc5}
  \begin{tabular}{@{}l l r r r r c r r r r@{}}
    \toprule
    & & \multicolumn{5}{c}{\textbf{Wall-clock time}}
    & \multicolumn{4}{c}{\textbf{Peak RSS (GB)}} \\
    \cmidrule(lr){3-7} \cmidrule(lr){8-11}
    \textbf{System} & \textbf{Dataset}
      & \textbf{T@25} & \textbf{T@50} & \textbf{T@75} & \textbf{T@100}
      & \textbf{Exp.}
      & \textbf{M@25} & \textbf{M@50} & \textbf{M@75} & \textbf{M@100} \\
    \midrule
    \multirow{4}{*}{\ArgusM}
      & Campus  & 25.1\,s & 38.1\,s & 50.6\,s & 2.4\,m  & 1.3 & 3.0  & 3.2  & 3.6  & 4.0  \\
      & \lanl   & 17.4\,m & 20.9\,m & 21.4\,m & 27.4\,m & 0.3 & 35.3 & 35.2 & 35.3 & 45.5 \\
      & \cicids & 2.4\,m  & 2.8\,m  & 3.0\,m  & 4.0\,m  & 0.4 & 7.8  & 7.7  & 7.7  & 7.4  \\
      & \optc   & 3.4\,m  & 8.0\,m  & 11.0\,m & 19.9\,m & 1.3 & 13.8 & 17.0 & 26.6 & 33.1 \\
    \midrule
    \multirow{4}{*}{\EulerM}
      & Campus  & 2.7\,m  & 3.5\,m  & 4.6\,m  & 5.6\,m  & 0.5 & 4.8 & 5.0 & 5.1 & 5.4 \\
      & \lanl   & 7.6\,m  & 8.7\,m  & 9.3\,m  & 17.0\,m & 0.6 & 3.7 & 4.3 & 6.3 & 7.8 \\
      & \cicids & 6.7\,m  & 7.4\,m  & 6.9\,m  & 8.8\,m  & 0.2 & 1.5 & 2.9 & 2.8 & 2.6 \\
      & \optc   & 5.8\,m  & 5.5\,m  & 6.1\,m  & 6.3\,m  & 0.1 & 0.5 & 0.6 & 0.7 & 0.8 \\
    \midrule
    \multirow{4}{*}{\PikachuM}
      & Campus  & 3.7\,m  & 7.9\,m  & 11.6\,m & 19.3\,m & 1.2 & 2.8  & 3.4  & 3.0  & 3.5  \\
      & \lanl   & 18.7\,m & 44.7\,m & 1.3\,h  & 4.7\,h  & 2.0 & 3.8  & 5.6  & 8.8  & 11.5 \\
      & \cicids & 3.5\,h  & 7.2\,h  & 13.0\,h & 17.3\,h & 1.2 & 15.0 & 31.5 & 56.6 & 67.9 \\
      & \optc   & 16.4\,m & 27.2\,m & 39.4\,m & 53.8\,m & 0.9 & 5.7  & 6.5  & 8.6  & 11.2 \\
    \midrule
    \multirow{4}{*}{\VgrnnM}
      & Campus  & 2.0\,m  & 2.3\,m  & 2.7\,m  & 2.7\,m  & 0.2 & 6.1 & 6.3  & 6.7 & 6.7  \\
      & \lanl   & 12.0\,m & 14.0\,m & 16.6\,m & 21.6\,m & 0.4 & 8.6 & 10.4 & 9.8 & 11.6 \\
      & \cicids & 30.6\,m & 32.0\,m & 31.7\,m & 33.0\,m & 0.1 & 7.8 & 7.7  & 7.8 & 7.8  \\
      & \optc   & 1.8\,m  & 2.7\,m  & 4.5\,m  & 5.6\,m  & 0.8 & 3.5 & 3.9  & 5.4 & 5.8  \\
    \midrule
    \multirow{2}{*}{\AnomalEM}
      & Campus  & 42.0\,m & 1.5\,h  & 2.3\,h & 3.3\,h & 1.1 & 10.6 & 20.6 & 30.5 & 40.6 \\
      & \cicids & 29.2\,m & 54.9\,m & 1.5\,h & 1.9\,h & 1.0 & 10.8 & 20.9 & 30.9 & 41.0 \\
    \bottomrule
  \end{tabular}
\end{table*}

Table~\ref{tab:app-gidslite-window} reports the same sweep for
\GidsLite (\S\ref{s:same-window}).  The two datasets where \GidsLite
ranks first by \ap are also the two where the window barely matters,
while on Campus and \lanl the \ap range moves substantially and on
\lanl the campaign exposure moves with it.  On \lanl the one-hour
window is also the best.  On the other three datasets the maximum lies
elsewhere, at 450\,s on Campus, 10\,min on \cicids, and 150\,min on
\optc.

\begin{table}[!t]
  \centering
  \footnotesize
  \setlength{\tabcolsep}{4pt}
  \renewcommand{\arraystretch}{1.10}
  \caption{\GidsLiteM sensitivity to snapshot-window size.  \ap swing
  is $(\max\ap-\min\ap)/\max\ap$ across the same window grid used in
  EG1.  \cdr swing is in percentage points.}
  \label{tab:app-gidslite-window}
  \begin{tabular}{@{}l c c c c@{}}
    \toprule
    \textbf{Dataset} & \textbf{\ap range} & \textbf{$\Delta$\ap}
      & \textbf{$\Delta$\cdr} & \textbf{Best window} \\
    \midrule
    Campus  & 0.278--0.354 & 21.3\% &  0.0 & 450\,s \\
    \lanl   & 0.112--0.510 & 78.1\% & 23.5 & 1\,h   \\
    \cicids & 0.980--0.988 &  0.8\% &  0.0 & 10\,m  \\
    \optc   & 0.976--0.993 &  1.8\% &  0.0 & 150\,m \\
  \bottomrule
  \end{tabular}
\end{table}

\subsection{EG2: Threshold-Selection Instability}
\label{app:sc2}

Table~\ref{tab:app-sc2} extends
Figure~\ref{fig:threshold-instability}(b) with the \fpr range, the
campaign-count range, and the count of metrics flagged unstable per
pair, all of which the figure does not display.  Only \Argus and
\Euler appear because the other three systems do not draw a random
validation sample for threshold calibration.
The unstable-count column shows the issue is rarely confined to one
metric: on the harder rows, the seed can move event detection, false
alerts, and campaign coverage together.  This is why the main text
treats threshold calibration as part of the protocol rather than a
cosmetic post-processing choice.

\begin{table}[!t]
  \centering
  \footnotesize
  \setlength{\tabcolsep}{2.5pt}
  \renewcommand{\arraystretch}{1.05}
  \caption{EG2 threshold-seed sensitivity.  Ranges are reported in
  percentage points across ten threshold-selection seeds.  The
  \emph{Unstable} column counts metrics whose seed-to-seed range
  exceeds reporting precision.}
  \label{tab:app-sc2}
  \begin{tabular}{@{}l l c r r r r c@{}}
    \toprule
    \textbf{System} & \textbf{Dataset} & \textbf{Seeds}
    & \textbf{$\Delta$\tpr} & \textbf{$\Delta$\fpr} & \textbf{$\Delta$\cdr}
    & \textbf{$\Delta$camp.} & \textbf{Unstable} \\
    \midrule
    \multirow{4}{*}{\ArgusM}
      & Campus  & 10 &  0.0 & 5.8 &  0.0 &  0.0 & 4 \\
      & \lanl   & 10 & 35.6 & 2.6 & 42.1 & 24.0 & 10 \\
      & \cicids & 10 &  9.3 & 2.8 & 30.8 &  8.0 & 8 \\
      & \optc   & 10 & 24.9 & 0.1 & 14.3 &  1.0 & 4 \\
    \midrule
    \multirow{4}{*}{\EulerM}
      & Campus  & 10 & 70.4 & 6.6 &  0.0 &  0.0 & 4 \\
      & \lanl   & 10 &  5.0 & 0.1 &  1.8 &  1.0 & 2 \\
      & \cicids & 10 &  0.0 & 0.0 &  0.0 &  0.0 & 1 \\
      & \optc   & 10 & 48.0 & 0.2 & 28.6 &  2.0 & 9 \\
    \bottomrule
  \end{tabular}
\end{table}

\subsection{EG3: Transductive Thresholding}
\label{app:sc3}

Table~\ref{tab:app-sc3} extends
Figure~\ref{fig:transductive-thresholding}(c) with \ap and \auc deltas,
which the figure does not display, and exact values for the
remaining post-threshold metrics.
The deltas should be read as operating-point shifts, not only accuracy
losses: an unchanged \tpr can come with a large \fpr increase.
The large \lanl shift is also shaped by the attack-vector
distribution in that dataset, not by protocol alone.  This is
the general case for inductive calibration: the split must be fixed
without knowing where attacks fall, so a protocol choice that looks
neutral decides what an evaluation can measure.

\begin{table}[!t]
  \centering
  \footnotesize
  \setlength{\tabcolsep}{4pt}
  \renewcommand{\arraystretch}{1.10}
  \caption{EG3 \PikachuM deltas.  Values report inductive minus
  transductive in percentage points.}
  \label{tab:app-sc3}
  \begin{tabular}{@{}l r r r r r r@{}}
    \toprule
    \textbf{Dataset}
      & \textbf{\ap} & \textbf{\auc} & \textbf{\tpr} & \textbf{\fpr}
      & \textbf{Prec.} & \textbf{$F_1$} \\
    \midrule
    Campus  & $-2.5$ &   0.3   & $-0.1$  &   0.7   & $-54.5$ & $-62.3$ \\
    \lanl   & $-9.1$ & $-71.1$ & $-66.4$ &  17.1   & $-9.3$  & $-16.3$ \\
    \cicids & $-1.0$ & $-6.2$  & $-83.6$ & $-16.2$ & $-68.5$ & $-75.3$ \\
    \optc   &   0.1  & $-13.5$ &   0.0   &  34.7   & $-0.3$  & $-0.2$  \\
    \bottomrule
  \end{tabular}
\end{table}

\begin{table*}[!t]
  \centering
  \footnotesize
  \setlength{\tabcolsep}{5pt}
  \renewcommand{\arraystretch}{1.05}
  \caption{EG7 option-level mean \ap.  Each row lists every option for
  one dataset and design dimension, sorted by mean \ap across the EG7
  configuration grid, with the best option listed first.}
  \label{tab:app-sc7}
  \begin{tabular}{@{}l l p{0.66\textwidth}@{}}
    \toprule
    \textbf{Dataset} & \textbf{Dimension} & \textbf{Options sorted by mean \ap} \\
    \midrule
    \multirow{4}{*}{Campus}
      & featurization     & identity\_structural (0.216), flow\_features (0.189), identity (0.178), structural (0.177) \\
      & graph encoder     & mpnn (0.269), graphsage (0.229), linear (0.191), gcn (0.134), none (0.127) \\
      & temporal encoder  & gru (0.198), none (0.198), lstm (0.184), gc\_lstm (0.180) \\
      & loss              & reconstruction (0.427), link\_prediction (0.122), contrastive (0.022) \\
    \midrule
    \multirow{4}{*}{\lanl}
      & featurization     & structural (0.239), flow\_features (0.180), identity\_structural (0.122), identity (0.099) \\
      & graph encoder     & none (0.254), graphsage (0.211), mpnn (0.153), linear (0.120), gcn (0.060) \\
      & temporal encoder  & none (0.210), gru (0.194), lstm (0.162), gc\_lstm (0.074) \\
      & loss              & reconstruction (0.269), link\_prediction (0.118), contrastive (0.092) \\
    \midrule
    \multirow{4}{*}{\cicids}
      & featurization     & structural (0.553), identity\_structural (0.546), identity (0.546), flow\_features (0.501) \\
      & graph encoder     & mpnn (0.601), none (0.554), graphsage (0.539), linear (0.501), gcn (0.487) \\
      & temporal encoder  & lstm (0.555), gru (0.546), gc\_lstm (0.532), none (0.514) \\
      & loss              & reconstruction (0.782), contrastive (0.480), link\_prediction (0.348) \\
    \midrule
    \multirow{4}{*}{\optc}
      & featurization     & identity (0.987), flow\_features (0.967), identity\_structural (0.962), structural (0.957) \\
      & graph encoder     & graphsage (0.971), mpnn (0.970), gcn (0.969), none (0.968), linear (0.963) \\
      & temporal encoder  & gc\_lstm (0.972), lstm (0.969), gru (0.966), none (0.966) \\
      & loss              & link\_prediction (0.989), contrastive (0.970), reconstruction (0.946) \\
    \bottomrule
  \end{tabular}
\end{table*}

\begin{table}[!t]
  \centering
  \scriptsize
  \setlength{\tabcolsep}{1.5pt}
  \renewcommand{\arraystretch}{1.20}
  \caption{EG8 native versus aligned preprocessing.  $\Delta$ columns
  report aligned minus native in percentage points.  Native (n) and
  aligned (a) are absolute scores.  Both arms use the per-dataset
  window fixed in EG8.}
  \label{tab:app-sc8}
  \begin{tabular}{@{}l l r r r r r r r r r@{}}
    \toprule
    & & \multicolumn{3}{c}{\textbf{\ap}} & \multicolumn{3}{c}{\textbf{\auc}}
    & \multicolumn{3}{c}{\textbf{\cdr}} \\
    \cmidrule(lr){3-5} \cmidrule(lr){6-8} \cmidrule(lr){9-11}
    \textbf{Sys.} & \textbf{Dataset}
      & \textbf{n} & \textbf{a} & \textbf{$\Delta$}
      & \textbf{n} & \textbf{a} & \textbf{$\Delta$}
      & \textbf{n} & \textbf{a} & \textbf{$\Delta$} \\
    \midrule
    \multirow{4}{*}{\ArgusM}
      & Campus  & $<.001$ & 0.017 &     1.6 & 0.124 & 0.559 &  43.5 & 0.000 & 0.000 &    0.0 \\
      & \lanl   & 0.105 & 0.104 & $-0.1$  & 0.996 & 0.818 & $-17.9$ & 0.649 & 0.105 & $-54.4$ \\
      & \cicids & $<.001$ & $<.001$ & 0.0  & 0.265 & 0.294 &   2.9 & 0.000 & 0.000 &    0.0 \\
      & \optc   & 0.334 & 0.521 &  18.7   & 0.979 & 0.973 & $-0.6$ & 1.000 & 1.000 &    0.0 \\
    \midrule
    \multirow{4}{*}{\EulerM}
      & Campus  & $<.001$ & 0.004 &     0.3 & 0.130 & 0.079 & $-5.1$ & 1.000 & 1.000 &    0.0 \\
      & \lanl   & 0.035 & 0.423 &  38.8   & 0.983 & 0.964 & $-1.9$ & 0.724 & 0.649 & $-7.5$ \\
      & \cicids & $<.001$ & $<.001$ & 0.0  & 0.248 & 0.316 &   6.8 & 0.000 & 0.000 &    0.0 \\
      & \optc   & 0.348 & 0.133 & $-21.5$ & 0.995 & 0.955 & $-4.0$ & 0.857 & 0.714 & $-14.3$ \\
    \midrule
    \multirow{4}{*}{\PikachuM}
      & Campus  & 0.896 & 0.915 &   1.9   & 0.989 & 0.990 &   0.1 & 1.000 & 1.000 &    0.0 \\
      & \lanl   & 0.091 & 0.059 & $-3.1$  & 0.711 & 0.619 & $-9.1$ & 0.596 & 0.702 &  10.5 \\
      & \cicids & 0.843 & 0.808 & $-3.5$  & 0.863 & 0.854 & $-1.0$ & 0.654 & 0.654 &    0.0 \\
      & \optc   & 0.988 & 0.987 & $-0.1$  & 0.994 & 0.995 &   0.1 & 1.000 & 1.000 &    0.0 \\
    \midrule
    \multirow{4}{*}{\VgrnnM}
      & Campus  & 0.015 & 0.445 &  43.0   & 0.909 & 0.960 &   5.1 & 0.000 & 0.000 &    0.0 \\
      & \lanl   & 0.002 & 0.620 &  61.8   & 0.974 & 0.969 & $-0.4$ & 0.737 & 0.912 &  17.5 \\
      & \cicids & 0.294 & 0.280 & $-1.4$  & 0.420 & 0.454 &   3.4 & 0.000 & 0.000 &    0.0 \\
      & \optc   & 0.503 & 0.935 &  43.2   & 0.989 & 0.974 & $-1.5$ & 1.000 & 1.000 &    0.0 \\
    \midrule
    \multirow{2}{*}{\AnomalEM}
      & Campus  & 0.024 & 0.048 &     2.4 & 0.950 & 0.500 & $-45.0$ & 1.000 & 1.000 &    0.0 \\
      & \cicids & 0.875 & 0.605 & $-27.0$ & 0.923 & 0.830 & $-9.3$ & 1.000 & 1.000 &    0.0 \\
    \bottomrule
  \end{tabular}
\end{table}

\subsection{EG4: Edge Features in \Argus}
\label{app:sc4}

Table~\ref{tab:app-sc4} extends Figure~\ref{fig:edge-feature-effect}
with the absolute \ap and \cdr values under both edge-feature
settings, which the figure shows only as deltas.  The cosine column
reports attack--benign overlap between the class means.  Because the
flow attributes are non-negative, that similarity is bounded well
above zero by construction and is weak evidence on its own; the
with/without deltas are what establish the feature path adds no
reliable signal.

\subsection{EG5: Cost and Scalability}
\label{app:sc5}

Table~\ref{tab:app-sc5} extends
Figure~\ref{fig:sc5-scaling-cost} from a worst-case summary to the
full per-pair sweep at all four input scales (25\%, 50\%, 75\%,
100\%).  The exponent column reports how wall-clock time grows between the
25\% and 100\% input scales, where 1.0 indicates linear growth
and values above 1.5 indicate super-linear growth.
Since the scaling exponents range from 0.1 to 2.0 across pairs,
reduced-scale runs cannot be extrapolated with one global factor.

\begin{table}[!t]
  \centering
  \footnotesize
  \setlength{\tabcolsep}{2pt}
  \renewcommand{\arraystretch}{1.05}
  \caption{EG4 \ArgusM edge-feature results.  Delta columns report
  with-feature minus without-feature in percentage points.  The cosine
  column reports the mean cosine similarity between attack and benign
  edge-feature vectors.}
  \label{tab:app-sc4}
  \begin{tabular}{@{}l r r r r r r r@{}}
    \toprule
    \textbf{Dataset}
      & \textbf{no-feat \ap} & \textbf{feat \ap} & \textbf{$\Delta$\ap}
      & \textbf{no-feat \cdr} & \textbf{feat \cdr} & \textbf{$\Delta$\cdr}
      & \textbf{Cos.} \\
    \midrule
    Campus  & 0.001 & 0.001 &     0.0  & 0.000 & 0.000 &     0.0  & 1.000 \\
    \lanl   & 0.014 & 0.105 &     9.1  & 0.719 & 0.649 & $-7.0$   & 1.000 \\
    \cicids & 0.002 & 0.002 & $-0.1$   & 0.654 & 0.269 & $-38.5$  & 1.000 \\
    \optc   & 0.760 & 0.033 & $-72.7$  & 1.000 & 0.000 & $-100.0$ & 0.979 \\
    \bottomrule
  \end{tabular}
\end{table}

\subsection{EG6: Covering-Edge Evasion}
\label{app:sc6}

\begin{table*}[!t]
  \centering
  \small
  {\renewcommand{\arraystretch}{1.25}%
  \setlength{\tabcolsep}{4pt}%
  \caption{Axis-by-axis comparison with Bilot
    \etal~\cite{bilot2025simpler} (PIDS) and Wang
    \etal~\cite{wang2025rr} (R+R), grouped by how each of our findings
    relates to the two prior studies.  The accompanying text
    (Appendix~\ref{app:prior-findings}) defines each category and
    discusses every row.}
  \label{tab:prior-findings}
  \begin{tabular}{@{}clm{3.9cm}m{3.5cm}m{5.1cm}@{}}
    \toprule
    \textbf{Relation} & \textbf{Axis} &
    \textbf{Bilot \etal~\cite{bilot2025simpler} (PIDS)} &
    \textbf{Wang \etal~\cite{wang2025rr} (R+R)} &
    \textbf{This work} \\
    \midrule
    \multirow{2}{*}[-5ex]{\RelNovel}
      & Graph boundaries & Kernel causality fixes the edges of the
        graph.  Streams are batched at fixed durations, but the
        duration is not studied. & The window is
        flagged as impactful, but measured one system at a
        time. & We sweep the window uniformly across systems and find
        a mean 38.3\% relative \ap swing (EG1). \\
      & Preprocessing & PIDS preprocessing follows shared DARPA
        Transparent Computing (TC) conventions~\cite{darpa}, so
        pipelines rarely differ. & Per-paper pipeline
        differences are noted, but their effect is not isolated. & A
        shared preprocessing contract isolates the effect: \Euler on
        \lanl moves from .035 to .423 \ap (EG8). \\
    \midrule
    \multirow{3}{*}[-7ex]{\RelInverts}
      & Edge features & In PIDS, symbolic features such as file paths
        separate attack from benign activity. & Over 93\% of Oilrig
        attack edges show feature similarity above 0.85 with benign
        edges. & On
        flows the feature path carries no reliable signal: it
        drops \Argus on \optc from 0.760 to 0.033 \ap (EG4). \\
      & Attacker control & PIDS detectors keep partial robustness
        under graph perturbation. & A single black-box evasion attack
        is run, on \lanl only. & Two
        inserted edges fully evade graph scoring on \lanl, and six of
        eight measurable detector--dataset pairs fall within twenty
        \mbox{inserted} edges (EG6). \\
      & Best components & For PIDS, one component recipe wins across
        all systems. & This axis is not studied. & For \gids
        the best component set is dataset dependent, with no universal
        recipe (EG7). \\
    \midrule
    \multirow{3}{*}[-7ex]{\RelConfirms}
      & Detection quality & A simple baseline is competitive with
        complex PIDS under a unified evaluation. & Published \gids
        results are fragile when re-run. & Both hold for \gids:
        published results are fragile when re-run, and a simple
        baseline is competitive once the protocol is matched. \\
      & Threshold hygiene & Threshold calibration is named as a PIDS
        evaluation shortcoming. & Per-system differences in
        thresholding protocol are noted. & Per detector--dataset
        pair, the seed moves \tpr by up to 70.4 percentage points and test-period
        calibration by up to 83.6 (EG2, EG3). \\
      & Cost & Scalability is priced for PIDS. &
        Runtime cost is not isolated. & We price every
        detector--dataset pair at four input scales, with peak memory
        reaching 67.9\,GB (EG5). \\
    \midrule
    \multirow{2}{*}[-5ex]{\RelAdapts}
      & Methodology & A staged decomposition of PIDS pipelines is
        introduced. & Each system is re-run end to
        end with its pipeline intact. & We adapt the staged
        decomposition to flows: six stages, per-component
        attribution. \\
      & Deployment & The real-time gap is noted and addressed with
        the streaming PIDS VELOX~\cite{bilot2025simpler}. & This axis
        is not studied. &
        Replaying archived executions, none of 18
        detector--dataset pairs is real-time ready (EG9). \\
    \bottomrule
  \end{tabular}}
\end{table*}

\begin{table}[!t]
  \centering
  \scriptsize
  \setlength{\tabcolsep}{2pt}
  \renewcommand{\arraystretch}{1.05}
  \caption{EG9 streaming readiness.  Replay columns mark whether the
  detector keeps up at the listed multiplier of the native ingest
  rate.  \emph{Win.} is the snapshot window in seconds, or a dash for
  static-graph systems; a dash under \emph{Max} means the detector
  keeps up at no tested rate.}
  \label{tab:app-sc9}
  \begin{tabular}{@{}l l c c c c c l l@{}}
    \toprule
    \textbf{Sys.} & \textbf{Dataset} & \textbf{Win.}
      & \textbf{1$\times$} & \textbf{2$\times$} & \textbf{5$\times$}
      & \textbf{Max} & \textbf{Verdict} & \textbf{Bottleneck} \\
    \midrule
    \multirow{4}{*}{\ArgusM}
      & Campus  & 150  & \cmark & \cmark & \cmark & 5 & near real-time & buffering \\
      & \lanl   & 3600 & \cmark & \cmark & \cmark & 5 & batch only     & buffering \\
      & \cicids & 900  & \cmark & \cmark & \cmark & 5 & batch only     & buffering \\
      & \optc   & 360  & \cmark & \cmark & \xmark & 2 & batch only     & buffering \\
    \midrule
    \multirow{4}{*}{\EulerM}
      & Campus  & 150  & \cmark & \cmark & \cmark & 5 & near real-time & buffering \\
      & \lanl   & 1800 & \cmark & \cmark & \cmark & 5 & batch only     & buffering \\
      & \cicids & 3000 & \cmark & \cmark & \cmark & 5 & batch only     & buffering \\
      & \optc   & 9000 & \cmark & \cmark & \cmark & 5 & batch only     & buffering \\
    \midrule
    \multirow{4}{*}{\PikachuM}
      & Campus  & 300  & \cmark & \cmark & \cmark & 5 & near real-time & buffering \\
      & \lanl   & 3600 & \cmark & \cmark & \cmark & 5 & batch only     & buffering \\
      & \cicids & 3000 & \cmark & \xmark & \xmark & 1 & batch only     & surge \\
      & \optc   & 3600 & \cmark & \cmark & \cmark & 5 & batch only     & buffering \\
    \midrule
    \multirow{4}{*}{\VgrnnM}
      & Campus  & 150  & \cmark & \cmark & \cmark & 5 & near real-time & buffering \\
      & \lanl   & 5400 & \cmark & \cmark & \cmark & 5 & batch only     & buffering \\
      & \cicids & 600  & \cmark & \xmark & \xmark & 1 & batch only     & surge \\
      & \optc   & 1800 & \cmark & \cmark & \cmark & 5 & batch only     & buffering \\
    \midrule
    \multirow{2}{*}{\AnomalEM}
      & Campus  & --   & \cmark & \cmark & \cmark & 5  & batch only & static rebuild \\
      & \cicids & --   & \xmark & \xmark & \xmark & -- & batch only & static rebuild \\
    \bottomrule
  \end{tabular}
\end{table}

\begin{table}[b]
  \centering
  \footnotesize
  \setlength{\tabcolsep}{1.5pt}
  \renewcommand{\arraystretch}{1.05}
  \caption{EG6 covering-edge evasion.  \emph{Cand.}\ counts the attack
  edges the attack processes and \emph{Flag.}\ the subset the detector
  flags.  ASR, in percent, is over the flagged subset only, so a row
  that flags nothing has no rate and is dashed throughout.
  \emph{Full} is the smallest budget at which no flagged edge
  survives, and a dash there means it did not happen by 50.  A
  further 126 candidates on \VgrnnPlain/\optc, and two to eight per
  LANL pair, were dropped for lack of an insertable covering edge.}
  \label{tab:app-sc6}
  \begin{tabular}{@{}l l c r r r r r r r c@{}}
    \toprule
    \textbf{Sys.} & \textbf{Dataset} & \textbf{Repro}
      & \textbf{Cand.} & \textbf{Flag.}
      & \textbf{@2} & \textbf{@5} & \textbf{@10}
      & \textbf{@20} & \textbf{@50} & \textbf{Full} \\
    \midrule
    \multirow{4}{*}{\ArgusM}
      & Campus  & cp & 355  & 0    &     -- &    -- &    -- &    -- &    -- & -- \\
      & \lanl   & rp & 440  & 352  & 100.0 & 100.0 & 100.0 & 100.0 & 100.0 & 2  \\
      & \cicids & ex & 215  & 145  &   0.0 &   0.0 &   0.0 &   0.0 &   0.0 & -- \\
      & \optc   & rp & 1080 & 1053 &  14.2 &  52.4 &  97.7 & 100.0 & 100.0 & 20 \\
    \midrule
    \multirow{4}{*}{\EulerM}
      & Campus  & cp & 355  & 0    &     -- &    -- &    -- &    -- &    -- & -- \\
      & \lanl   & rp & 466  & 407  &  94.8 & 100.0 & 100.0 & 100.0 & 100.0 & 5  \\
      & \cicids & ex & 76   & 0    &     -- &    -- &    -- &    -- &    -- & -- \\
      & \optc   & rp & 1080 & 1059 &  31.7 &  66.7 & 100.0 & 100.0 & 100.0 & 10 \\
    \midrule
    \multirow{4}{*}{\VgrnnM}
      & Campus  & cp & 355  & 0    &     -- &    -- &    -- &    -- &    -- & -- \\
      & \lanl   & rp & 428  & 371  & 100.0 & 100.0 & 100.0 & 100.0 & 100.0 & 2  \\
      & \cicids & ex & 119  & 3    & 100.0 & 100.0 & 100.0 & 100.0 & 100.0 & 2  \\
      & \optc   & rp & 96   & 96   &  12.5 &  25.0 &  62.5 &  89.6 &  93.8 & -- \\
    \bottomrule
  \end{tabular}
\end{table}

Table~\ref{tab:app-sc6} extends Figure~\ref{fig:sc6-perturbation} with
the reproduction level (\emph{rp} for paper repro, \emph{cp} for code
port, \emph{ex} for protocol extension), the candidate counts, and the
numeric ASR at every budget.  The \emph{Flag.}\ column is the attack surface that exists at all:
where it is zero the detector misses the campaign before an adversary
acts, and where it is large, low-budget success recurs across
reproductions and extensions.

\subsection{EG7: Component Sweep}
\label{app:sc7}

Table~\ref{tab:app-sc7} extends
Table~\ref{tab:sc7-frameworked-ablation} from one best option
per dimension to the full sorted list of every option, with its mean
\ap over every grid configuration that uses it.  This shows the gap
between the winner and the alternatives, including dimensions where
multiple options tie within reporting precision.  In several datasets,
removing graph or temporal encoders remains competitive with heavier
choices, so attribution should come from controlled component sweeps
rather than from the presence of a \gnn block.

\subsection{EG8: Native Versus Aligned Preprocessing}
\label{app:sc8}

Table~\ref{tab:app-sc8} extends
Figure~\ref{fig:sc8-preprocessing-impact} from its six pairs to all 18
and adds the \auc and \cdr metrics that the figure does not display.
The same row shows whether a preprocessing change moves all three
metrics in the same direction, or only one: some rows improve \ap
while campaign coverage stays flat or moves the other way, so
preprocessing can change separability and the operational alert set.

\subsection{EG9: Streaming Readiness}
\label{app:sc9}

Table~\ref{tab:app-sc9} extends Table~\ref{tab:sc9-streaming-readiness}
from an aggregate count of replay-job verdicts to the per-pair
detail.  Each row records whether the detector keeps up at
1$\times$, 2$\times$, and 5$\times$ the native ingest rate, the
snapshot window in seconds, and the dominant bottleneck.  The
\emph{Bottleneck} column distinguishes the buffering delay (waiting
for a snapshot to close) from surge backlog and from the static-graph
rebuild path used by \AnomalE.  Many rows keep up even at 5$\times$,
but the first alert is still delayed by snapshot buffering or
static-graph rebuild, so batch throughput does not make
them real-time detectors.

\section{Comparison with Prior Studies}
\label{app:prior-findings}

Two prior studies frame this comparison.  Wang
\etal~\cite{wang2025rr} re-ran five published \gids end to end,
keeping each system's pipeline intact, and showed that reported
detection metrics change by up to 40\% under undocumented assumptions,
preprocessing discrepancies, and hyperparameter and
threshold-calibration sensitivity.  Bilot
\etal~\cite{bilot2025simpler} studied provenance-based intrusion
detection (PIDS), a neighboring problem in which the graph is derived
from kernel audit causality rather than network traffic.  Under a
unified evaluation they showed that a simple neural baseline matches
or outperforms much more complex graph-based systems, and they
identified nine recurring shortcomings of PIDS evaluation.  The two
settings differ in their data model: the edges of a PIDS graph and its
symbolic features, such as file paths and command lines, are given by
the operating system, whereas a \gids must construct its graph from a
continuous flow stream by choosing a snapshot window, a preprocessing
pipeline, and statistical edge attributes.  This difference decides
which findings can transfer between the two settings and which cannot.

Table~\ref{tab:prior-findings} classifies each of our findings by its
relation to these two studies, using four categories.  \emph{Novel}
marks a measurement that neither prior study provides.  \emph{Inverts}
marks an axis that Bilot \etal also examine for PIDS, where our
network-flow answer is the opposite of theirs.  \emph{Confirms} marks
a prior conclusion that holds under our matched protocol.
\emph{Adapts} marks an axis or methodology that we carry
over from prior work rather than claim as new.

The two novel rows exist because the \gids data model turns graph
construction itself into a free parameter.  In PIDS the edges of the
provenance graph are given by kernel causality, and systems batch the
event stream at fixed durations, such as fifteen-minute
intervals.  The duration sets how much of a fixed graph is scored at
once, and Bilot \etal encounter it only on the deployment axis, as a
real-time limitation, not as a sensitivity to sweep.  In \gids the
snapshot window instead decides which flows aggregate into which
graph, so it shapes the graph the detector sees.  Wang \etal flag it
as impactful but measure one system at a time.  Sweeping it uniformly
across systems and datasets yields a mean 38.3\% relative \ap swing
(EG1).  Similarly, PIDS preprocessing has converged on the shared
conventions of the DARPA Transparent Computing (TC)
engagements~\cite{darpa}, while every \gids paper ships its own
pipeline.  Wang
\etal note these differences without isolating their effect.  Imposing
one shared preprocessing contract isolates it, moving \Euler on \lanl
from .035 to .423 \ap (EG8).

The three inverted rows follow from the feature gap between the two
data models.  In PIDS, symbolic features carry the separating signal,
which is why Bilot \etal find that feature-rich pipelines detect well,
that PIDS detectors keep partial robustness under graph perturbation,
and that one component recipe wins across systems.  On network flows
each of these answers flips.  The statistical edge attributes of \gids
are near-collinear across attack and benign traffic, so enabling the
feature path drops \Argus on \optc from 0.760 to 0.033 \ap (EG4).  The
graph is built from attacker-observable traffic, so two inserted edges
fully evade graph scoring on \lanl and six of the eight measurable
pairs fall within a budget of twenty inserted edges (EG6).  The best component set
changes from dataset to dataset, with no universal recipe (EG7).

The confirming rows are the two headline conclusions of the prior
studies, and both survive the move to network flows.  Wang \etal
report that published \gids results are fragile when re-run,
and our sweeps agree: the seed alone moves \tpr by up to 70.4 percentage
points and test-period calibration by up to 83.6 percentage points
(EG2, EG3).  Bilot
\etal found that a simple baseline is competitive for PIDS once the
evaluation is matched, and \GidsLite repeats this for \gids, ranking
first by \ap on two of the four datasets.  The shared cost-hygiene
concern becomes a per-pair measurement: every
detector--dataset pair is priced at four input scales (EG5).

The adapted rows are debts rather than discoveries.  The staged
decomposition is the methodology of Bilot \etal, redefined for flows
and extended to per-component attribution.  The deployment axis they
address for PIDS with the streaming system
VELOX~\cite{bilot2025simpler} becomes our replay measurement, which
finds none of 18 detector--dataset pairs real-time ready (EG9).

\end{document}